\documentclass[oldversion]{aa}  
\usepackage{ulem}
\usepackage{graphicx}
\usepackage{txfonts}
\usepackage{multirow}

\begin{document}

   \title{X-ray to NIR emission from AA~Tauri during the dim state}
\subtitle{Occultation of the inner disk
and gas-to-dust ratio of the absorber}
\titlerunning{New FUV and X-ray observations of AA~Tau}

   \author{P. C. Schneider \inst{1,2}
          \and
          K. France \inst{3,7}
          \and
          H. M. G\"unther\inst{4}
          \and
          G. J. Herczeg \inst{5}
         \and
          J. Robrade \inst{2}
          \and          
          J. Bouvier \inst{6}
          \and
          M. McJunkin \inst{3}
          \and
          J. H. M. M. Schmitt \inst{2}
          }

   \institute{
   European Space Research and Technology Centre (ESA/ESTEC), Keplerlaan 1, 2201 AZ Noordwijk, The Netherlands
      \email{christian.schneider@esa.int}
   \and
   Hamburger Sternwarte, Gojenbergsweg 112,
              Hamburg, 21029 Germany
         \and
             Center for Astrophysics and Space Astronomy, University of Colorado, 389 UCB, Boulder, CO 80309, USA
        \and             
             Harvard-Smithsonian Center for Astrophysics,
             60 Garden Street, Cambridge, MA 02139, USA
         \and
           The Kavli Institute for Astronomy and Astrophysics, Peking University, Yi He Yuan Lu 5, Hai Dian Qu, Beijing 100871, China
         \and Univ. Grenoble Alpes, IPAG, F-38000 Grenoble, France\\
         CNRS, IPAG, F-38000 Grenoble, France
          \and Laboratory for Atmospheric and Space Physics,
University of Colorado, 392 UCB, Boulder, CO 80309}

   \date{Received 24/12/2014; accepted 20/08/2015.}

\abstract{
      AA Tau is a well-studied, nearby classical T~Tauri star, which  
      is viewed almost edge-on. A warp in its inner disk 
      periodically eclipses the central star, causing a clear 
      modulation of its optical light curve. The system underwent 
      a major dimming event beginning in 2011 caused by an extra
      absorber, which is most likely associated with additional 
      disk material in the line of sight toward the central 
      source. We present new XMM-Newton X-ray, Hubble Space 
      Telescope FUV, and ground based optical and 
      near-infrared data of the system obtained in 2013 during the 
      long-lasting dim phase. The  line width decrease 
      of the fluorescent H$_2$ disk emission shows that the 
      extra absorber is located at
      $r>1$\,au. Comparison of X-ray absorption ($N_H$) with dust 
      extinction ($A_V$), as derived from measurements obtained one inner 
      disk orbit (eight days) after the X-ray measurement, indicates that the 
      gas-to-dust ratio as probed by the $N_H$ to $A_V$ ratio of the extra absorber is 
      compatible with the ISM ratio.   Combining both 
      results suggests that the extra absorber, i.e., material 
      at $r>1\,$au, has no significant gas excess in contrast 
      to the elevated gas-to-dust ratio previously derived for 
      material in the inner region ($\lesssim0.1\,$au).
  }

   \keywords{Stars: individual: AA Tau,  Stars: low-mass, stars: pre-main sequence,
   stars: variables: T Tauri, Herbig Ae/Be,  X-rays: stars,  Ultraviolet: stars
               }

   \maketitle

\section{Introduction}
Disks around protostars play a key role in star formation and influence
many features of the final stellar system. For example,
most of the stellar mass is acquired by disk accretion, 
the rotation period is set by magnetic interaction with the
disk, and planets  form within the disk.
These so-called accretion or protoplanetary disks contain 
approximately $10^{-3}$ to $10^{-1}\,M_\odot$
and disperse within a typical time span of a few Myrs 
\citep[see review by ][]{Alexander_2013}, which also sets the 
time scale for planet formation. Within this time,      
disks undergo large structural changes. In particular, grains grow to 
larger sizes, and
larger grains are thought to settle in the disk midplane leaving
a gas rich disk atmosphere behind \citep[see review by][]{Williams_2011}.

Gas comprises the majority of the mass in a protoplanetary disk, with 
only a small fraction contributed by dust. The canonical 
gas-to-dust ratio of 100:1 found in the interstellar medium is often
also assumed for protoplanetary disks. Therefore, the gas content 
controls essential transport processes within the disk like
angular momentum redistribution and dust grain motion.
Gas affects grain growth through the coupling of
gas and dust dynamics \citep{Weidenschilling_1977, Takeuchi_2001},
as well as the thermal and chemical balance of the disk 
\citep[e.g.,][]{Woitke_2009}. 

Measuring the disk's dust content is relatively straightforward
because the thermal dust emission is readily observable
at infrared and radio wavelengths. However, 
measuring the gas content of protoplanetary disks is more challenging. 
Observational studies often rely on gas emission lines, but their 
interpretation is highly model dependent. For example, 
\citet{Tilling_2012} modeled the disk around the Herbig Ae star HD~163296 
and found gas-to-dust ratios between 9:1 and 100:1 depending on their model 
assumptions. A specific problem is that the bright CO emission can become 
optically thick, so that only part of the gas mass is traced 
\citep[e.g.,][]{Hughes_2008} and that
CO freezes-out on grains at low ($\sim20$\,K) temperatures in 
the disk midplane. In addition, gas and dust emission are
not necessarily co-spatial, thus introducing additional uncertainties
into the local gas-to-dust ratio.

Transmission spectroscopy is a different approach to studying 
the gas content of (nearly) edge-on disks \citep[e.g.,][]{Lahuis_2006,Rettig_2006, Schneider_2010, Horne_2012,
France_2012_AA_Tau, McJunkin_2013}.
In particular, X-ray absorption provides a complementary view
of the gas-to-dust ratio, because it is sensitive to the total amount of 
material in the line of sight, i.e., it is largely dominated by
gas in most scenarios. Comparing X-ray and dust extinction thus provides
an estimate of the gas-to-dust ratio for a given line of sight. This 
estimate depends on different assumptions than gas emission line 
studies, e.g., the total X-ray absorption
is dominated by He and heavier elements (e.g., oxygen, carbon, iron)
and not by hydrogen, which represents the main component of protoplanetary
disks. Therefore, transforming the X-ray absorption to an equivalent hydrogen 
column density implicitly assumes a particular abundance pattern. 

Here, we use X-ray absorption spectroscopy to study the disk
around the classical T~Tauri star (CTTS) AA~Tau. 
A so-far unique dimming event of the system allows us to 
study new regions of the disk using transmission spectroscopy.
Combined with new measurements of the dust extinction from
optical and near-IR (NIR) photometry and new Hubble Space Telescope (HST)
far-ultraviolet (FUV) spectroscopy of disk emission lines, we
derive the disk's gas-to-dust ratio and provide an estimate for
the disk region traced by this measurement.

Our paper is structured as follows. 
We start with a brief description of the AA~Tau
system in the next section (sect.~\ref{sect:AA_Tau}) and
describe observations and data analysis in 
sect.~\ref{sect:data}. Using the X-ray data, we measure essentially the 
gas column toward the stellar source in sect.~\ref{sect:Xray}.
Using optical and near-IR data, we derive the dust content of the 
line of sight  in
sect.~\ref{sect:extinction}. We then compare the dust extinction
with the absorbing gas column derived from the X-ray data to study the gas-to-dust 
ratio (sect.~\ref{sect:cf}).
In sect.~\ref{sect:location} we study fluorescent H$_2$
line emission from the disk surface to constrain the location of
the extra absorber. We close with a summary and discussion in 
sect.~\ref{sect:summary}. 

\begin{table*}[t]
\begin{minipage}[h]{0.99\textwidth}
\centering
\renewcommand{\footnoterule}{}
  \caption{Analyzed observations \label{tab:obs} }
  \begin{tabular}{c c c c c c c } \hline \hline
  Observing date & Observatory & Instrument &  Filter/grating & Exp. time & Approx. wavelengths& Obs. ID.\\     
  \hline
   \multirow{3}*{2013 Aug 15} & \multirow{3}*{XMM-Newton}& EPIC & Medium & 34\,ks & 3 -- 30\,\AA & \multirow{3}*{0727960501} \\
                              &                          & \multirow{2}*{OM} &  V, B & 6.0\,ks each & 551, 445\,nm\\
                              &                          &                   & UVW1, UVW2 & 6.5, 6.7\,ks & $\approx$300, 212\,nm & \\[0.3cm]

  \multirow{2}*{2011 Jan 06} & \multirow{2}*{HST} & \multirow{2}*{COS} & G160M &  4.2\,ks & \multirow{2}*{1133 -- 1795\,\AA}  & \multirow{2}*{11616} \\
              &  &   & G130M &  5.7\,ks & &  \\[0.3cm]
  2013 Feb 06 & \multirow{2}*{HST} & \multirow{2}*{COS} & G160M &  8.7\,ks &  \multirow{2}*{1065 -- 1790\,\AA\footnote{With $\sim$10 \AA\ and 25 \AA\ gaps around 1217 and 1375\,\AA, respectively}}& \multirow{2}*{12876}\\
  2013 Feb 05 &  &  & G130M & 18.2\,ks & & \\[0.3cm]
  2013 Aug 23 & UKIRT & WFCAM & J, H, K & $5\times1$\,s & 1.25, 1.65, 2.2\,$\mu$m & U/13A/H28B \\[0.3cm]
  2013 Oct 16 & \multirow{3}*{CAHA} & \multirow{3}*{Omega 2000} & \multirow{3}*{J, H, K} & \multirow{3}*{$4\times8$\,s} & \multirow{3}*{1.25, 1.65, 2.2\,$\mu$m} & \multirow{3}*{DDT}  \\
  2013 Oct 17 & &  &  &  &  & \\
  2013 Oct 21 & &  &  &  &  & \\
  \hline
  \end{tabular}
  \end{minipage}
\end{table*}

\section{AA~Tau: An almost edge-on CTTS \label{sect:AA_Tau}}

AA~Tau is a well-studied low-mass CTTS 
in the Taurus star forming region at a distance of 140\,pc 
\citep{Kenyon_1994}.
Its spectral type is M0.6 -- K7 
with an $A_V$ of $0.4$ -- $0.8$ \citep[][resp.]{Herczeg_2014, Bouvier_1999}.
We adapt $M_\star \approx 0.8\,M_\odot$ as estimated by \citet{Bouvier_1999}.
Disk accretion rates between $10^{-9}$  and  
$10^{-8}\,M_\odot$\,yr$^{-1}$ have been derived  
\citep[e.g.,][]{Basri_1989,Valenti_1993,Gullbring_1998,Bouvier_2003,Bouvier_2013}.
The  AA~Tau system is seen close to edge-on ($i\sim75^\circ$) so 
that the  light passes through the upper layers of its 
protoplanetary disk \citep[e.g.,][]{Bouvier_2003, Cox_2013}.

AA~Tau is the prototype of stars showing periodic features in 
their optical light curves caused by variable line of sight 
extinction. These so-called AA~Tau-like variables show a stable 
maximum brightness and pronounced, periodic minima (with 
$\Delta V\approx1-2$\,mag). These minima are most likely caused 
by a  disk warp that periodically (partially) 
occults the star. For AA~Tau, the typical eclipse depth 
during the last 24\,years is approximately 1.0 to 1.6\,mag while 
there are also short periods in time when no or only little 
additional extinction is seen \citep{Bouvier_2003}. Assuming 
Keplerian 
rotation, the observed periodicity of about eight days indicates 
that  the warp is located at 0.07\,au, i.e., 
very close to the inner edge of the dust disk. We refer to 
this feature as the ``inner warp'' in the following.

The periodicity of the extinction allows one to 
separate the features of the inner warp from those of other 
absorbers along the line of sight. 
\citet{Bouvier_2007} suggest that the grains in the inner warp 
are larger than in the ISM. 
\citet{Schmitt_2007} show from their analysis of a series of 
XMM-Newton observations that the line of sight passes through 
material with a gas-to-dust ratio about ten times higher than
in the ISM and associate this gas excess with a dust-depleted 
outflow or dust-free accretion streams \citep[see also][for a 
detailed analysis of these X-ray data in the multi-wavelengths 
context]{Grosso_2007}.

The gas content along the line of sight was also studied 
using high-resolution FUV observations. HST COS observations 
have shown that the 
neutral atomic hydrogen column density toward AA~Tau is 
$4-5\times10^{20}\,$cm$^{-2}$ \citep[][]{France_2012_AA_Tau, McJunkin_2013b}. 
This is lower than the minimum expected from 
the optical extinction ($A_V=0.4 - 0.8$) for an ISM-like gas-to-dust ratio
and at least a  factor of 18 below the column density derived from 
X-ray observations. This discrepancy is probably caused by
a high molecular fraction within the circumstellar material, 
leaving little atomic hydrogen in the disk atmosphere, 
as shown by \citet{France_2014_CO} for RW~Aur (a CTTS in a 
comparable evolutionary stage as AA~Tau). 
Molecular absorption lines have also been measured in the FUV with
H$_{2}$ absorption  tracing the hot ($T\sim2500$\,K) molecular content 
\citep[$N_{H_2}=8^{+22}_{-4}\times10^{17}$\,cm$^{-2}$, see][]{France_2012_AA_Tau}
and CO absorption bands tracing the warm ($T\sim500$\,K) molecular material
 \citep[$N_{CO} = 3^{+7}_{-2}\times10^{17}$\,cm$^{-2}$, see ][]{France_2012_AA_Tau,McJunkin_2013}.
For the canonical H$_2$ to CO ratio \citep[$\sim 10^4$, e.g., ][]{Lacy_1994} one would therefore 
expect an absorbing column density of $\gtrsim10^{22}\,$cm$^{-2}$ 
in agreement with the X-ray data ($N_H\sim1-2\times10^{22}\,$cm$^{-2}$).

\begin{figure*}[t!]
\centering
\includegraphics[width=0.95\textwidth, angle=0]{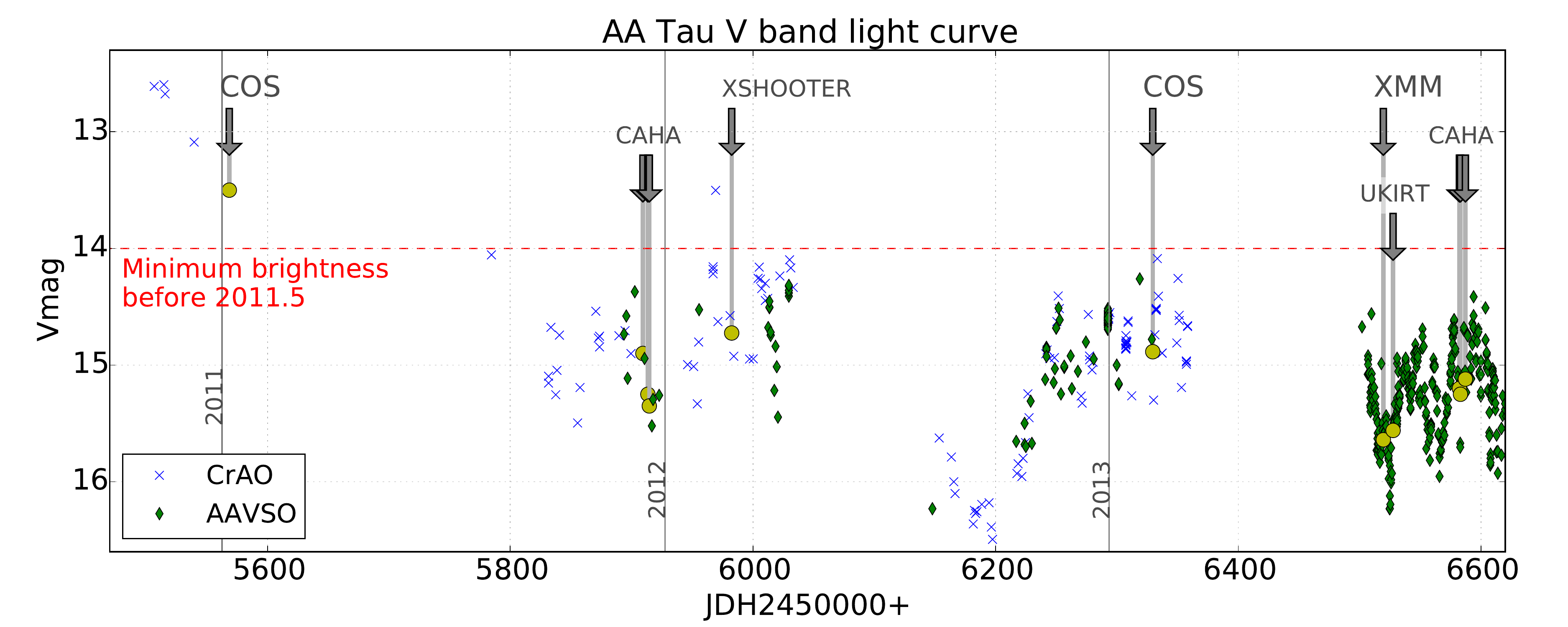}
\caption{V-band light curve of AA~Tau in heliocentric Julian date. The important 
observations are labeled. The yellow circles indicate the estimated brightness 
during these observations. The red, dashed line indicates the lowest state prior
to the dimming.\label{fig:lc} }
\end{figure*}

Recently, \citet{Bouvier_2013} have reported that AA~Tau experienced an 
unexpected drop in its optical brightness by \hbox{about 2 mag} in
the V band toward the end of 2011. Figure~\ref{fig:lc} shows the
optical light curve with the dates of the analyzed observations marked. 
In the following, we denote the new state as the ``dim state'' and the
previous state at higher optical luminosity as the ``bright state''. 
The most likely explanation for the dimming is that a part of the 
disk located farther out than the inner warp is eclipsing the system, 
because emission from the star is obscured, but 
the spatially extended jet emission has not changed. 
We term the origin of the increased extinction the ``extra absorber''
in the following.
From time-scale arguments, \citet{Bouvier_2013} estimated a radius 
of about eight au for the obscuring material 
and their VLT-Xshooter observations during the dim state show no change in 
mass accretion rate onto AA~Tau.   The wavelength dependence of 
the additional reddening suggests that scattered light contributes a 
large fraction of the blue-optical flux  during the dim state
contrary to the bright state for which \citet{Bouvier_1999} 
estimated a negligible scattering fraction. Thus, scattering of
optical photons likely contributes
also to our new 2013 data.

\section{Observations and data analysis \label{sect:data}}
Table~\ref{tab:obs} lists the new X-ray, FUV, optical and NIR data 
obtained during the dim state of AA~Tau.
Specifically, we obtained new XMM-Newton data to 
measure the gas content of the line of sight
(sect. \ref{sect:Xray_description}) and new \hbox{(near-)}
simultaneous optical (NIR) 
data  that trace the dust content (sect.~\ref{sect:NIR_data}).
In addition, we use new HST COS data described in sect.~\ref{sect:HST_COS} 
to study H$_2$ emission lines originating in the disk surface
to constrain the location of the extra absorber.

\subsection{XMM-Newton X-ray data \label{sect:Xray_description}}

The new XMM-Newton observations of AA~Tau  during the dim state
were performed on 2013 August 15 for 34\,ks.
We 
concentrate on the EPIC CCD spectra obtained with MOS1, MOS2, and 
pn \citep[for instrument descriptions see ][]{Turner_2001,Strueder_2001}. 
XMM-Newton data reduction has been performed using SAS 
version~12\footnote{\texttt{http://xmm.esac.esa.int/sas/}}. Periods of 
enhanced particle background have been 
excluded from the spectral analysis. Filtered exposure times are 
28.6\,ks for pn and 33.3\,ks for each MOS detector. 
There are no strong flares in the X-ray light curve and we fit 
the entire observation with a single set of parameters.
The three EPIC spectra were fitted simultaneously using XSPEC \citep{XSPEC}. 
Therefore, the fits are dominated by the pn data as 
the pn detector provides the highest count number.
We used an absorbed APEC model \citep{APEC} for
the spectral modeling (XSPEC notation: \texttt{phabs * vapec}) 
as in \citet{Schmitt_2007} and adopt their  plasma abundances for 
consistency. The \texttt{phabs} model uses the absorption cross sections 
of \citet{phabs} and we used elemental abundances for the absorber according 
to \citet{Anders_1989}. The equivalent hydrogen column density ($N_H$)
depends on the abundances assumed for the absorber; a lower metalicity implies 
a higher $N_H$.  Unless noted otherwise, we use the \citet{Anders_1989} abundances
for the absorber to be consistent with previous X-ray studies of AA~Tau.
For the relevant photon energy range ($E_{phot}>1.0$\,keV)
differences between this absorption model and the \texttt{wabs} model used by
\citet{Grosso_2007} are negligible.
Errors indicating 90\,\% confidence ranges are provided for the 
X-ray data.

To compare the dim state with the previous bright state, we 
use archival XMM-Newton data from an observing campaign
that obtained eight snapshots in 2003 covering two rotation periods 
of the inner disk warp;
these data have previously been published by \citet{Schmitt_2007} and 
\citet{Grosso_2007}.

\setlength{\tabcolsep}{4pt}
\begin{table}[t]
\caption{Near-IR magnitudes of AA~Tau. \label{tab:NIR_data} }
  \begin{tabular}{l c l r r l l}\hline\hline
  Date & Observ. & V\tablefootmark{1} & J & H & K & Ref.\\
  \hline
  1995 Nov 16\tablefootmark{2} & CAMILA & 13.9 & 9.99 & 9.14 & 8.55 & (1)\\
  1995 Nov 21\tablefootmark{3} & CAMILA & 12.6 & 9.30 & 8.6 & 8.11 & (1)\\ 
  1997 Nov 30 & 2MASS & -- & 9.43 & 8.55 & 8.05\tablefootmark{4} & (2)\\
  2011 Dec 13 & \multirow{3}*{CAHA} & 14.9 & 10.46 & 9.20 & 8.46 & \multirow{3}*{(3)}\\
  2011 Dec 17 &  & 15.3 & 10.61 & 9.27 & 8.43 \\
  2011 Dec 19 &  & 15.4& 10.53 & 9.21 & 8.45 \\[0.2cm]
  2013 Aug 23 & UKIRT & 15.54 & 11.13 & 10.13 & 9.70 & (4) \\[0.2cm]
  2013 Oct 16 & \multirow{3}*{CAHA} & 15.17 & 11.00 & 9.63 & 8.81 & \multirow{3}*{(4)}\\
  2013 Oct 17 & & 15.15 & 10.85 & 9.52 & 8.65\\
  2013 Oct 21 & & 15.05 & 10.59 & 9.32 & 8.58\\
  \hline
  \end{tabular}
  \tablefoot{\tablefoottext{1}{Estimated from the data points close 
             in time to the NIR measurements.}
             \tablefoottext{2}{Inner warp in front of AA~Tau.}
             \tablefoottext{3}{Inner warp behind AA~Tau.}
             \tablefoottext{4}{The 2MASS filter is K$_s$ and not K.}
  }
            
  \tablebib{
(1)~\citet{Bouvier_1999}; (2)~\citet{Skrutskie_2006}; (3)~\citet{Bouvier_2013}; (4)~This work
}
\end{table}
\renewcommand{\arraystretch}{1}

\subsection{Optical and near-IR data around the XMM-Newton observation 
\label{sect:NIR_data} \label{sect:Xphase}}
The evolution of the optical and NIR magnitudes
of AA~Tau depends on the dust content of the line of sight
and we use new optical/NIR data to estimate the dust 
extinction during the XMM-Newton observation.

During the X-ray observation, we obtained simultaneous V, B,
UVW1, and UVW2 images with the optical monitor (OM) on-board of XMM-Newton.
The pipeline source detection reports magnitudes of $15.63\pm0.01$, 
$17.16\pm0.02$, and $17.38\pm0.08$
for the V, B, and UVW1 filters using the data from Exp.\,IDs S006 -- S008.
The UVW1 filter is centered on the bright Mg~{\sc ii} doublet that is
known to be jet related, i.e., we do not consider it as a good
tracer of photometric emission, but include it in Fig.~\ref{fig:sed}
for completeness.
The B and V magnitudes are within the range
of previous measurements during the dim state \citep{Bouvier_2013} so that
the V magnitude is 1.6 to 3.1\,mag below the values during the 2003 XMM-Newton
observing campaign.
No source is found by the detection algorithm in the UVW2 exposure, 
but extracting the counts in a circle with a radius of 4.8 arcsec (10\,pixel) 
the UVW2 exposure ($\lambda_{cen} = 2120\,$\AA) shows slight excess
counts above the expected background at the 
position of AA~Tau significant at about 90\,\% confidence level\footnote{The 
statistical error is given by the chance to find the number of detected 
source photons by a random background fluctuation.}.
The approximate magnitude is $19.4^{+2.2}_{-0.4}$.
Compared to the 2003 XMM-Newton campaign when the OM was also operated with
the UVW2 filter, this is nominally 2.6 -- 3.6\,mag darker
(observed magnitude range in 2003: \hbox{$\sim$ 15.8--16.8\,mag}).

In addition, AA~Tau has been monitored around the XMM-Newton observation by 
the Crimean Astrophysical Observatory (CrAO, K. Grankin, priv. communication)
and by astronomers around the world organized through the AAVSO (Henden, A.A., 2013, 
Observations from the AAVSO International Database, http://www.aavso.org).
These data provide the frame
to study periodic features in the light curve as 
expected from the inner warp.

Near-IR data were obtained by UKIRT\footnote{See \texttt{www.ukirt.hawaii.edu}} 
and by Omega2000\footnote{\texttt{www.caha.es/CAHA/Instruments/O2000/index.html}} 
at the Calar-Alto 3.5\,m telescope (CAHA). 
Pipeline reduction has been used for the UKIRT data and
custom python routines for the Omega2000 data including
sky subtraction and flat-fielding. 
Table~\ref{tab:NIR_data} lists the resulting near-IR magnitudes of 
AA~Tau together with estimates of the V magnitude during the
near-IR observations from nearby optical data points. 
This gives a J magnitude 
around 11 at V=15.6. The corresponding H and K band magnitudes are 
about 10 and 9.7, respectively.

For the XMM-Newton observation, we estimate the near-IR magnitudes from
the nearest measurements as no  
simultaneous observations exist, i.e., from the UKIRT data 
that were obtained at approximately similar optical brightness. 
Fortunately, the UKIRT data have been obtained about eight days after
the XMM-Newton observation so that both observations trace the same
phase with respect to the inner disk warp. 

\begin{figure}[t!]
\centering
\includegraphics[width=0.48\textwidth, angle=0]{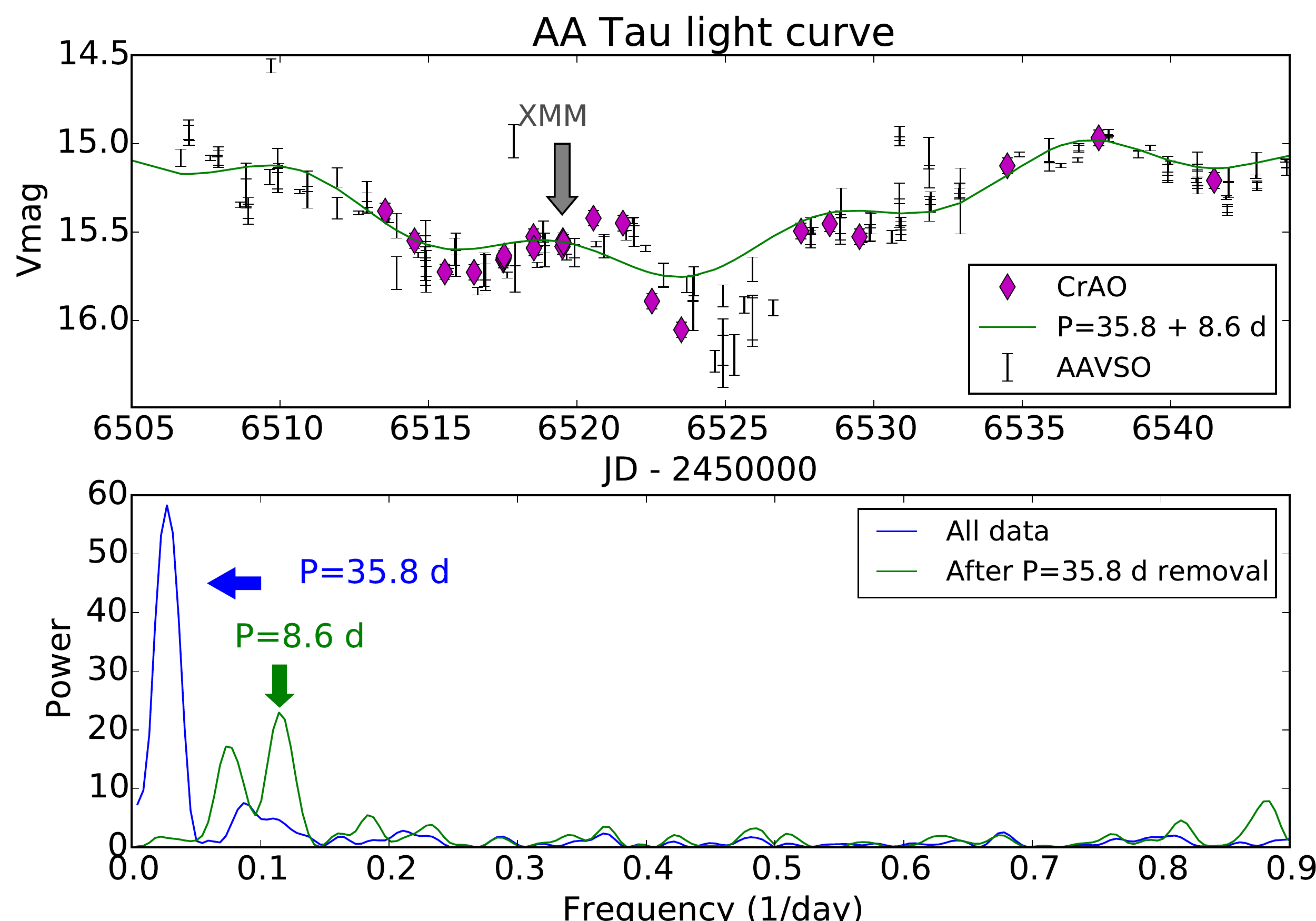}
\caption{{\bf Top}: Close-up of the optical light curve around the XMM-Newton
observation with sinusoidal models. {\bf Bottom: }Lomb-Scargle diagram with 
the peaks marked. \label{fig:lc_xmm} }
\end{figure}

Figure~\ref{fig:lc_xmm} (top) shows a close-up of the optical light curve 
around the XMM-Newton observation. AA~Tau showed periodic variations in
its light curve in the bright state and we search for similar variability
in the new optical data (starting MJD 2456501) using Lomb-Scargle periodograms 
\hbox{(Fig.~\ref{fig:lc_xmm} bottom}, phased light curves are provided in 
Fig.~\ref{fig:timing_new}). The light curve shows variation 
on a time scale much longer than the original eight day period as indicated by 
the 36\,day peak in the periodogram. We remove this long term trend from the 
light curve and find a modulation with an 8.6\,day period, which is very 
close to the original 8.19\,day periodicity
\citep[][statistical false alarm probability is $10^{-9}$]{Artemenko_2012}. 
The good match between both periods shows that variability caused by the inner 
warp is still present in the optical light curve. The derived period is 
probably influenced by a longer, possibly irregular pattern. 
We think that a comparable additional modulation of the light curve might
be responsible for the fact that the optical data presented by 
\citet{Bouvier_2013} did not show a significant periodicity if analyzed 
as one single data set, but that a significant periodic signal was 
found in a specific three month time interval with an amplitude of 0.7\,mag in 
the V band. In fact, the modulation amplitude is expected to vary as the 
ratio between direct and scattered light changes with absolute
magnitude. 
Inspection of Fig.~\ref{fig:lc_xmm} 
reveals that the XMM-Newton observation was obtained when the majority 
of the inner warp was not in the line of sight towards AA~Tau.  
The statistical uncertainty of the phase is small, so that we think
that a reasonable estimate of the phase uncertainty is rather given
by the phase difference that results from leaving the period as a free fit 
parameter ($P=8.6$\,day model) and a model with the period set to 
its previous value ($P=8.19$\,days), i.e., we consider a phase 
uncertainty of 0.2 or 1.6 days as a reasonable estimate.

\subsection{FUV spectra from HST COS \label{sect:HST_COS}}
Two HST COS FUV observations of AA~Tau exist. The 2011 data, 
published by \citet{France_2012_AA_Tau},  were obtained during the 
bright state while the 2013 FUV data 
were obtained during the dim state.
Thus, changes in the disk emission between both 
epochs allow us to put constraints
on the location of the extra absorber.

The wavelength ranges covered by the COS observations are 
listed in Tab.~\ref{tab:obs}
and multiple FP offsets have been used to mitigate the effects 
of fixed pattern noise.
The COS FUV data of 
both epochs were processed as described in detail by 
\citet{France_2014}. 
The one-dimensional spectra produced by the COS calibration pipeline, 
CALCOS, were aligned and co-added using custom software procedures 
described by \citet{Danforth_2010}. The analysis of the H$_{2}$ lines 
follows the method described in \citet{France_2012_H2}. Line profiles 
were obtained by 
weighting the individual central wavelength settings with their 
respective exposure time for all line fitting applications. 
A new observing mode (\texttt{cenwave}=$\lambda\,1222$)
was used for the short wavelength
observations in 2013 to extend the available spectrum down to 
1065~\AA{} which, 
however, causes the loss of any information on the Ly$\alpha$ 
emission line so that changes in the atomic and molecular 
hydrogen absorption between both epochs must remain unexplored.
 
Part of the description of the FUV data can be found in the
appendix. While these spectra are interesting in their own right, 
they are compatible with the results of the main text and do 
not provide additional constrains. In particular, these results 
relate to the evolution of the atomic emission lines, the CO emission 
lines, as well as the CO absorption seen against the continuum. 
We refer the interested reader to the respective appendix sections 
(App.~\ref{sect:App_cont} to 
\ref{sect:CO_abso}).

\begin{figure}[t!]
\centering
\vspace*{-0.5cm}\includegraphics[width=0.48\textwidth, angle=0]{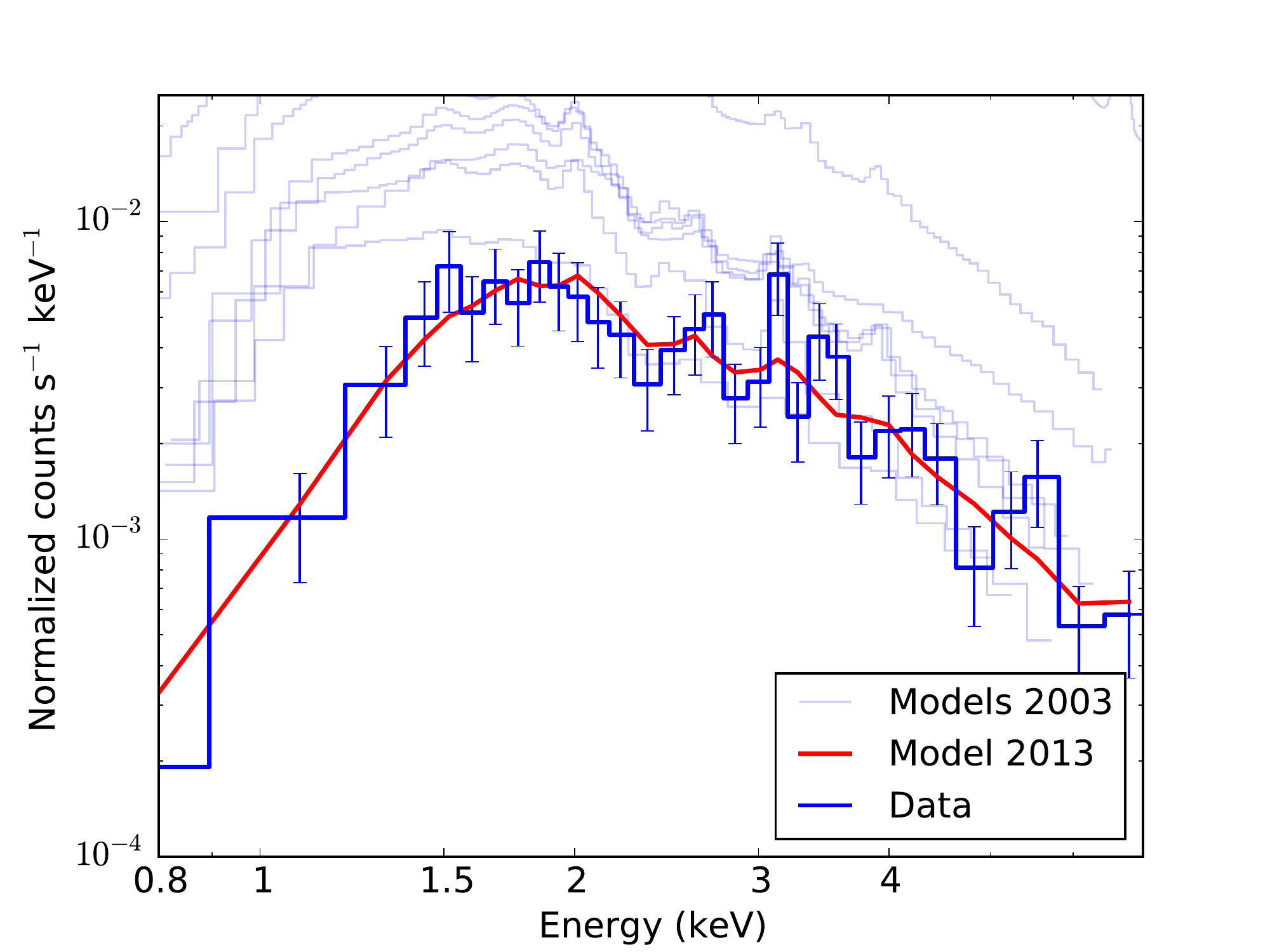}
\caption{Comparison of the X-ray spectrum during the dim state with models for
the archival X-ray spectra obtained during the bright state (Models 2003). 
The pn-spectrum binned to 15\,counts per bin is shown as Data. 
\label{fig:X-ray} }
\end{figure}

\section{X-ray absorbing column density towards AA~Tau: The line of sight's 
gas content \label{sect:Xray}}

X-ray absorption is sensitive to the gas content of the line of sight
and, thus, provides complementary information about the extra absorber
compared to optical/NIR data which are essentially sensitive to the 
dust content. We derive the X-ray absorption parameterized by $N_H$ 
from our new XMM-Newton data through modeling the coronal emission of 
AA~Tau with one absorbed thermal emission component. 
Table~\ref{tab:Xray_results} lists the fit results. 
The coronal properties (temperature and emission measure) of AA~Tau
derived from the new 2013 data
are within the range of values observed during the 2003 XMM-Newton 
campaign. This indicates that the stellar source remains rather
unaffected by the process(es) that changed the extinction.
Figure~\ref{fig:X-ray} compares the new X-ray observation with
models for the archival XMM-Newton observations from 2003. 
The column density best fitting the 2013 X-ray data 
($N_H=1.9\times10^{22}$\,cm$^{-2}$) is slightly higher than the 
highest value during the 2003 observing campaign when the X-ray 
absorption varied between $0.9$ and $1.7\times10^{22}$\,cm$^{-2}$. 
Assuming the \citet{aspl} abundances for the absorbing
material, the equivalent hydrogen column density is
$N_H=2.85_{-0.45}^{+0.47}\times10^{22}\,$cm$^{-2}$.

Although scattering must be considered for optical photons, X-ray
scattering is regularly ignored in the context of CTTSs. Nevertheless, we
verified that scattering is unlikely to affect the X-ray spectrum of AA~Tau. 
At X-ray energies Thomson scattering (energy independent 
and only moderately angle dependent) and Rayleigh Gans scattering (energy 
dependence $E_{phot}^{-2}$ and forward scattering) can contribute for gas 
and grain scattering, respectively. For our estimate, we assume that the 
scattering screen extends $\pi$\,sr (a quarter of the sky; 
a very conservative estimate) and $A_V=5$ (for dust scattering) or 
$N_H=2\times 10^{22}\,$cm$^{-2}$ (for Thomson scattering). Further assuming 
no absorption along the scattering light path, the maximum 
scattering fraction is $\lesssim10\,\%$ at 1\,keV (about the softest
energies in the observed X-ray spectrum). Therefore, we ignore scattering 
when interpreting the observed X-ray spectrum of AA~Tau \citep[see also][ for
a discussion of X-ray scattering in the context of CTTSs]{Bally_2003}.

\begin{table}[t]
\begin{center}
\renewcommand{\footnoterule}{}
\caption{X-ray results. Fluxes are provided for the 0.2 -- 10.0\,keV energy range. 
Values for the 2003 data are taken from \citet{Schmitt_2007}. Values in 
brackets pertain to the 2003 observation
with strongly enhanced X-ray emission. This observation stands out from the
rest of the observations and might not be representative of the X-ray 
emission of AA~Tau. \label{tab:Xray_results} }
  \begin{tabular}{l c c l}\hline\hline
  Parameter & 2013 &  2003 & Unit\\
  \hline
  Count rate\tablefootmark{1} &  21.6 & 19 -- 147 (800) & counts\,ks$^{-1}$ \\[0.2cm]
  $N_H$ & $1.9\pm{0.3}$ & $0.9^{-0.4}_{+0.5}$ -- $1.7^{+0.3}_{-0.3}$ & $10^{22}\,$cm$^{-2}$\\[0.2cm]
  k$T$ & $3.1_{-0.7}^{+1.1}$ & $1.8_{-0.4}^{+0.8}$ -- $3.9_{-0.5}^{+0.7}$ & keV\\[0.2cm]
  $EM$ & $4.0_{-0.8}^{+1.0}$ & $3.5_{-1.1}^{+1.5}$ -- $21.6_{-2.0}^{+2.6}$  & $10^{52}$\,cm$^{-3}$\\
       &                     &  ($141.4_{-5.3}^{+5.4}$) & \\[0.2cm]
  $L_{X\,obs.}$ & 2.2 & 1.5 -- 16.4 (96.4) & $10^{29}$\,erg\,s$^{-1}$\\[0.2cm]
  $L_{X\, emitted}$ & 5.6 & 4.0 -- 31.8 (208) & $10^{29}$\,erg\,s$^{-1}$\\[0.2cm]
  $\chi^2$/d. o. f. & 33.1 / 51 & 23/21 - 139/134 (835/801) & --\\
  \hline
  \end{tabular}
  \tablefoot{\tablefoottext{1}{Summed EPIC count rate}}
\end{center}
\end{table}

To estimate the absorption caused by the extra absorber, we compare the 
new $N_H$ measurement with the values observed in 2003. 
Using the full range of the 2003 $N_H$ values, i.e.,
assuming no dependence of $N_H$ with inner disk phase 
\citep[as described by][]{Grosso_2007}, we find 
$N_H^{\text{extra}} = 0.2 \dots 1.0\times10^{22}\,$cm$^{-2}$ with statistical
errors of $0.4-0.5\times10^{22}$\,cm$^{-2}$.
Excluding the 2003 values closest to mid eclipse that have the highest
column density ($N_H=1.70$ and $1.73\times10^{22}$\,cm$^{-2}$), the range 
reduces to  $N_H^{\text{extra}} = 0.5 \dots 1.0\times10^{22}$\,cm$^{-2}$ with 
similar uncertainties. Thus, the extra column density that can be associated
with the extra absorber is below $1.5\times10^{22}$\,cm$^{-2}$.

\section{Dust extinction towards the stellar source \label{sect:extinction}}
We now analyze the dust extinction caused by the extra 
absorber during the XMM-Newton observation for comparison 
with the X-ray derived column density. We denote this extra
absorption with $\Delta A_V$ in the following.

\begin{figure}[t!]
\centering
\includegraphics[width=0.47\textwidth, angle=0]{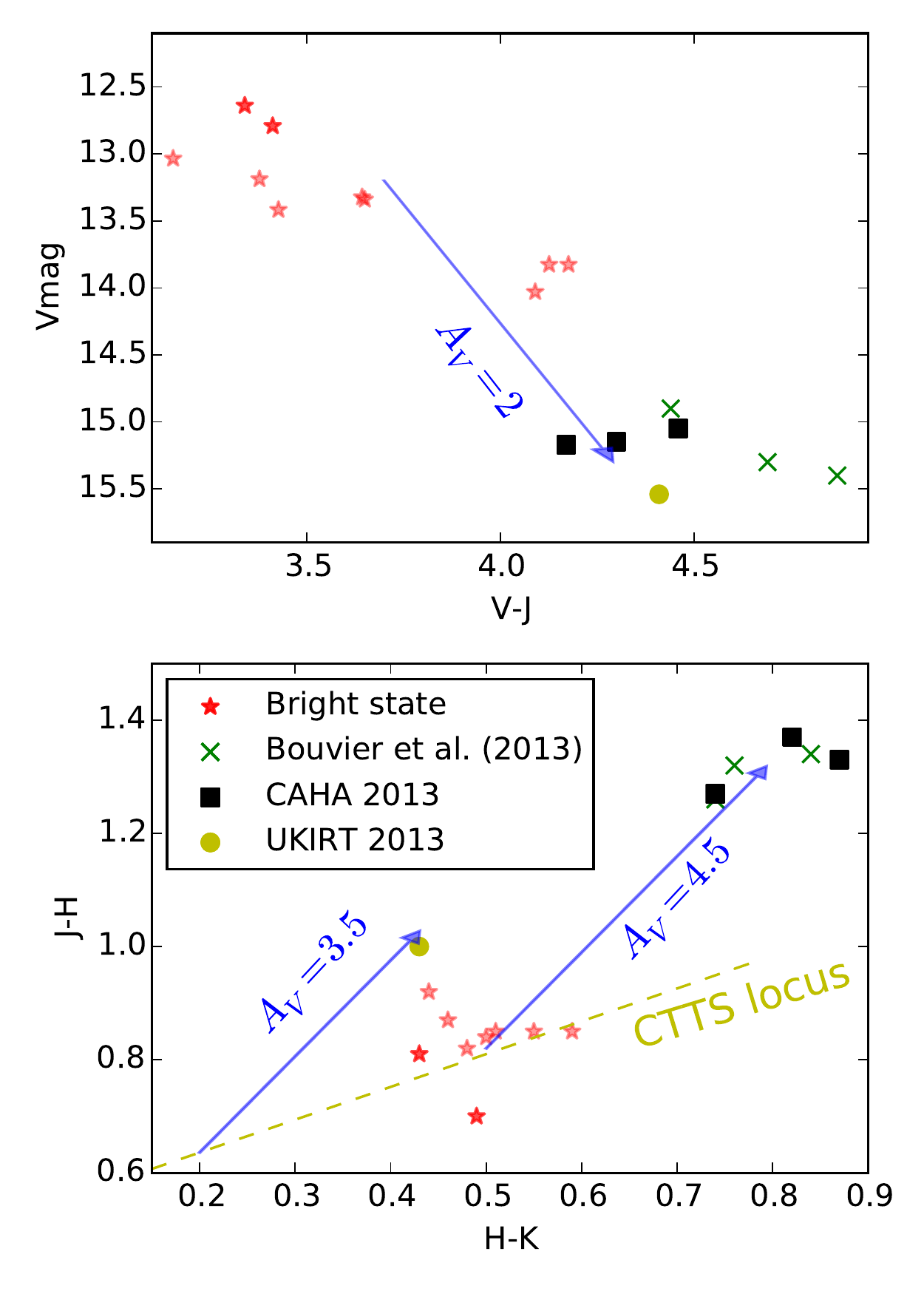}
\caption{Color-magnitude (top) and color-color diagrams (bottom) of AA~Tau.
The CTTSs locus is from \citet{Meyer_1997}. Extinction vectors from
\citet{Rieke_1985}. The darker red stars for the bright state indicate
measurements with the inner warp behind AA~Tau.
\label{fig:color_mag} }
\end{figure}

Circumstellar material around CTTSs might contain processed
dust so that the extinction law potentially deviates from its 
standard ISM form \citep[e.g., see review by ][]{Testi_2014}. In
addition, stellar photons might be scattered
within the circumstellar environment erroneously suggesting a
low dust extinction, i.e., $\Delta V \neq \Delta A_V$.
Scattering is strongly wavelength dependent ($\sim\lambda^{-4}$).
Therefore, we use the evolution of the NIR magnitudes 
to estimate the dust extinction during the X-ray observation.
Specifically, we deredden the NIR colors  to the CTTS locus
using the \citet{Rieke_1985} extinction law (see Fig.~\ref{fig:color_mag} 
bottom). This gives
$A_V\approx3.5$ for the UKIRT observation and $A_V\approx4.5$ for 
the CAHA data, i.e., within 0.5\,mag of the results by \citet{Bouvier_2013}
who estimated $A_V\approx4$. 
For the bright state, most of
the data points are distributed around the CTTS locus implying 
little to no NIR extinction (especially when considering that we
are interested in the values for the situation when the inner warp 
is behind AA~Tau). Thus,
the extra absorber is
responsible for most of the extinction implying $\Delta A_V\approx4\pm0.5$
during the X-ray observation. As a check, we use prior
knowledge on the J magnitudes of AA~Tau during the bright state
and attribute the drop in J 
magnitude to the extra absorber, i.e.,  we assume $\Delta J = A_J$ 
and transfer $A_J$ to $A_V$ using the \citet{Rieke_1985} extinction law.
We do not consider the H and K magnitudes, because they are likely 
affected by disk emission as 
suggested by the NIR color-color diagram (Fig.~\ref{fig:color_mag})
and by the magnitude spread, 
which is smallest in J. Comparing the 
bright state with the inner warp behind AA~Tau 
to the J magnitudes from 2013 (Tab.~\ref{tab:NIR_data}),
we find $\Delta J = 1.3 - 1.8\,$mag, which transforms to 
$A_V = 4.6 - 6.4$. This is slightly higher than the first method. 
The highest value ($A_V=6.4$) pertains to the UKIRT 
data for which dereddening to the CTTS locus gives $A_V\approx3.5$.
Therefore, we extend the confidence range on $A_V$ to 
higher values and adapt $A_V=4.0_{-0.5}^{+1.0}$ for 
the dust extinction during the dim state.

\begin{figure*}[t!]
\centering
\includegraphics[width=0.9\textwidth, angle=0]{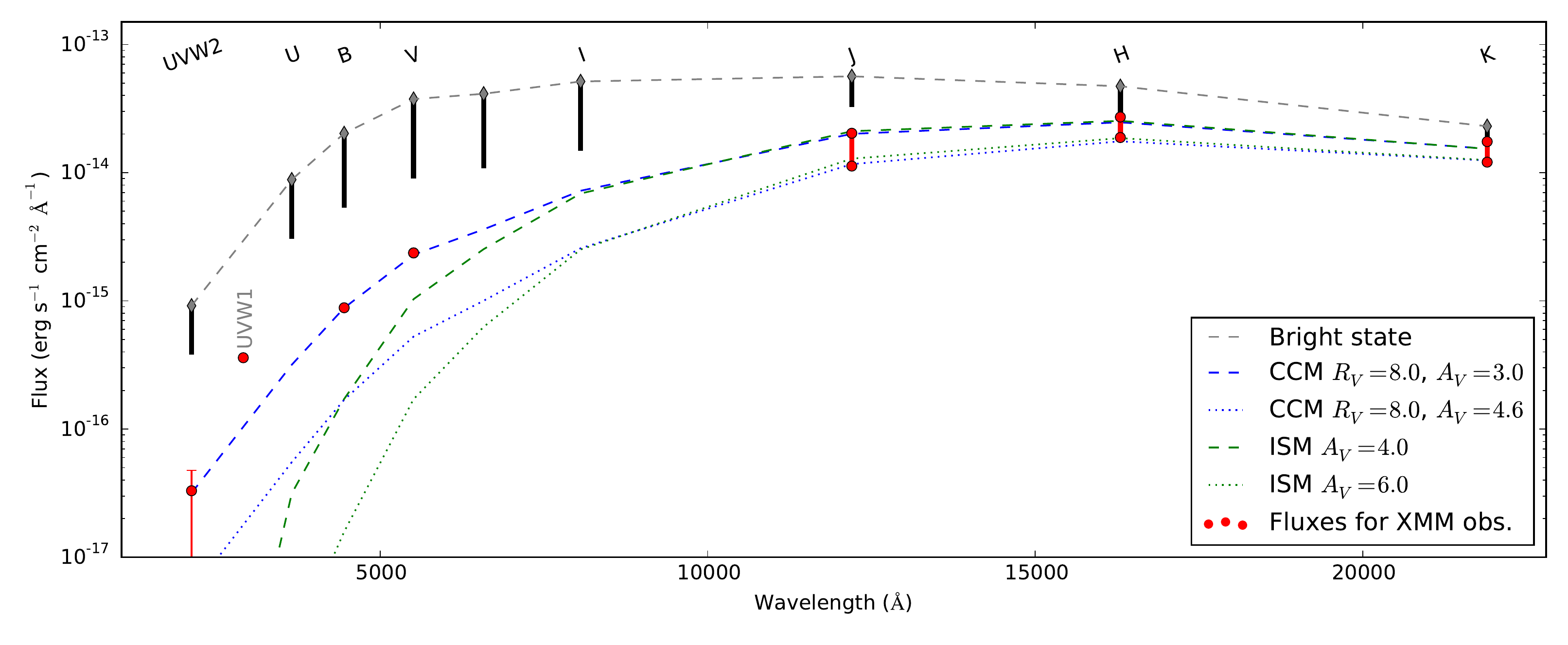}
\caption{Fluxes for the AA~Tau system during the bright and dim state. 
During the bright state the observed photospheric emission was modulated by the inner warp
and the gray bars indicate the ranges for each filter.
The fluxes during the XMM-Newton observation are shown as red circles. The red bars indicate the range of NIR magnitudes observed during the
dim state. The gray, dashed line indicates the uneclipsed SED during the bright state for comparison with the dim state during 
the XMM-Newton observation. Finally, the blue and green lines (dashed and dotted) show the expected flux distribution taking 
the gray line as the reference assuming different extinction laws \citep[CCM refers to the][extinction law]{CCM}.\label{fig:sed} }
\end{figure*}

Figure~\ref{fig:sed} shows the range of fluxes observed 
during the bright state and around the 2013 XMM-Newton observation. 
The broad wavelength coverage of our observations
allows us to investigate if
scattering contributes to the observed flux (scenario I)
or if a deviation from the ISM extinction law parameterized
by changing $R_V$ can explain the
observed flux evolution without scattering (scenario II).
This $R_V$ parameterization might not be entirely correct for the 
circumstellar environment of CTTSs, but nevertheless gives 
an impression of the effect and is used for illustrative 
purposes here. Specifically, we consider
the extinction laws of \citet{Fitzpatrick_1999} and \citet{CCM} 
setting $R_V$ to higher values than the canonical 3.1 to mimic
grain growth in an approximate way. We do not draw
conclusions based on the numerical value of $R_V$ since we are
only interested in the impact of a different grain size 
distribution on the extinction curve and in particular on the 
NIR extinction.

Dust extinction increases towards shorter wavelengths so that
FUV data can provide 
strong constrains on the extinction. However, intrinsic variability 
as well as an outflow contribution hamper direct interpretation of 
the AA~Tau data. App.~\ref{sect:app_FUV_lines} describes that
the observed evolution is compatible with the two following
scenarios providing no additional discriminating power.

\subsection{Scenario I: Scattering \label{sect:dust_scattering}}

\citet{Bouvier_2013}
suggest that scattering causes the stellar V magnitude  
to decrease less strongly than expected from their near-IR
photometry. With $\Delta A_V \sim 4$ from the NIR magnitudes
and $\Delta V$=2.5, we find a similar behavior.
In this scenario, the majority of the observed optical photons 
do not travel through the inner warp but 
are scattered. This causes the amplitude of the  eight day light curve 
modulation to decrease. A scattering fraction of a few
percent causes a constant flux of V$\sim$15.6. The remaining modulation
is caused by the direct light passing through the inner warp.
Assuming $\Delta A_V\approx4.5$ suggested by the CAHA data, the 
optical brightness variation decreases from the value observed during 
the bright state ($\Delta V \approx 1.5$) to $\Delta V \approx 0.2$,
i.e., reasonably close to the observed value of 0.15.

\subsection{Scenario II: Dust evolution \label{sect:scattering}}
Figure~\ref{fig:sed} shows
that the \citet{CCM} extinction law with $\Delta A_V=3.0$ and $R_V=8.0$  
approximately reproduces the observed flux evolution between the 
bright and dim states. 
Redenning the bright state fluxes  matches
the optical data points and the resulting near-IR magnitudes
are close to the brightest values observed during the dim state. 
However, the magnitudes observed around the 2013 XMM-Newton 
observation are rather high compared to the full range of 
observed magnitudes 
(see sect.~\ref{sect:NIR_data}). 
Using even higher $R_V$ values cannot bring these NIR magnitudes 
in agreement with the optical/NUV magnitudes. This tension might
be caused by time variability between the XMM-Newton and NIR
observations. However, the observed decrease of the eight day optical 
light modulation due to the inner warp is not expected within this scenario 
and would require changes within the AA~Tau system in addition
to an extra absorber.
Applying the \citet{Fitzpatrick_1999} extinction law
results in steeper (redder) spectral energy distributions
than the \citet{CCM} models and fits the data less well.

In summary, the scattering scenario is compatible with the available 
data while the grain growth scenario requires additional model 
modifications to reconcile the data. Therefore, we consider scattering
more likely. Its strong wavelengths dependence 
allows us to estimate  $\Delta A_V =4.0_{-0.5}^{+1.0}$ during the 
XMM-Newton observation in 2013.

\section{Comparing $N_H$ and $A_V$ \label{sect:cf}}
Using the gas absorption characterized by $N_H$, measured from 
the X-ray data, and the dust extinction from the optical/NIR data described 
by $A_V$, we evaluate the gas-to-dust ratio of the extra absorber
and compare it to previous measurements during the bright state.
We use \hbox{$A_V\approx4.0_{-0.5}^{+1.0}$} for the dust extinction caused by the extra 
absorber 
(see sect.~\ref{sect:extinction}) and the full range gas column densities
found in sect.~\ref{sect:Xray} 
\hbox{($N_H^{\text{extra}} = 0.2 \dots 1.0\pm0.5\times10^{22}\,$cm$^{-2}$)}.
This results in \hbox{$N_H = 0.5\dots2.5\pm1.4\times10^{21}$\,cm$^{-2}$\,$A_V^{-1}$},
which is is compatible with the ISM value of 
\hbox{$N_H = 1.8\times10^{21}$\,cm$^{-2}$\,$A_V^{-1}$}
\citep[e.g.,][]{Predehl_1995}. This ratio is lower than
the about ten times elevated gas content found during the 2003 
XMM-Newton campaign and suggests that the extra absorber
is less gas-rich than the material in the line of sight of the 
2003 observations.

Within the uncertainty margin a gas free absorber as well as a slightly gas-rich 
composition are possible.
Within about 0.1\,au, one 
finds that $N_H/A_V$ is of the order of $10^{22}$\,cm$^{-2}$\,$A_V^{-1}$ 
(see sect.~\ref{sect:AA_Tau}), i.e., about an order of magnitude
above the ratio for the extra absorber. This shows that the extra absorption 
is not caused by material with the same gas-to-dust ratio as found
in the inner region of the AA~Tau system.

\section{H$_2$ disk emission: The location of the extra absorber 
\label{sect:location}}
\citet{Bouvier_2013} suggest that the extra absorber is located
several au from AA~Tau. This is compatible 
with the optical light curve showing less pronounced modulation 
as well as with the lower $N_H/A_V$ value, which suggests that
the extra absorber's composition is different from the material
within 0.1\,au. 
Here, we use the evolution of the molecular hydrogen 
emission thought to originate in the upper disk layers to put further 
constrains on the location of the extra absorber. 

H$_2$ at temperatures around $2-3\times10^3$\,K is fluorescently 
pumped by stellar Ly$\alpha$ photons 
and decays primarily via discrete transitions in the FUV.
The spatial origin of the H$_2$ emission 
lines is imprinted in their observed velocity structure since
the dominant broadening mechanism of these lines is the Doppler shift
due to Keplerian rotation around the central star 
\citep{France_2012_H2}. Therefore, comparing the velocity structure
of the H$_2$ emission during the dim and the bright state
constrains the disk location obscured by the extra absorber.

\begin{figure}
\centering
\includegraphics[width=0.49\textwidth, angle=0]{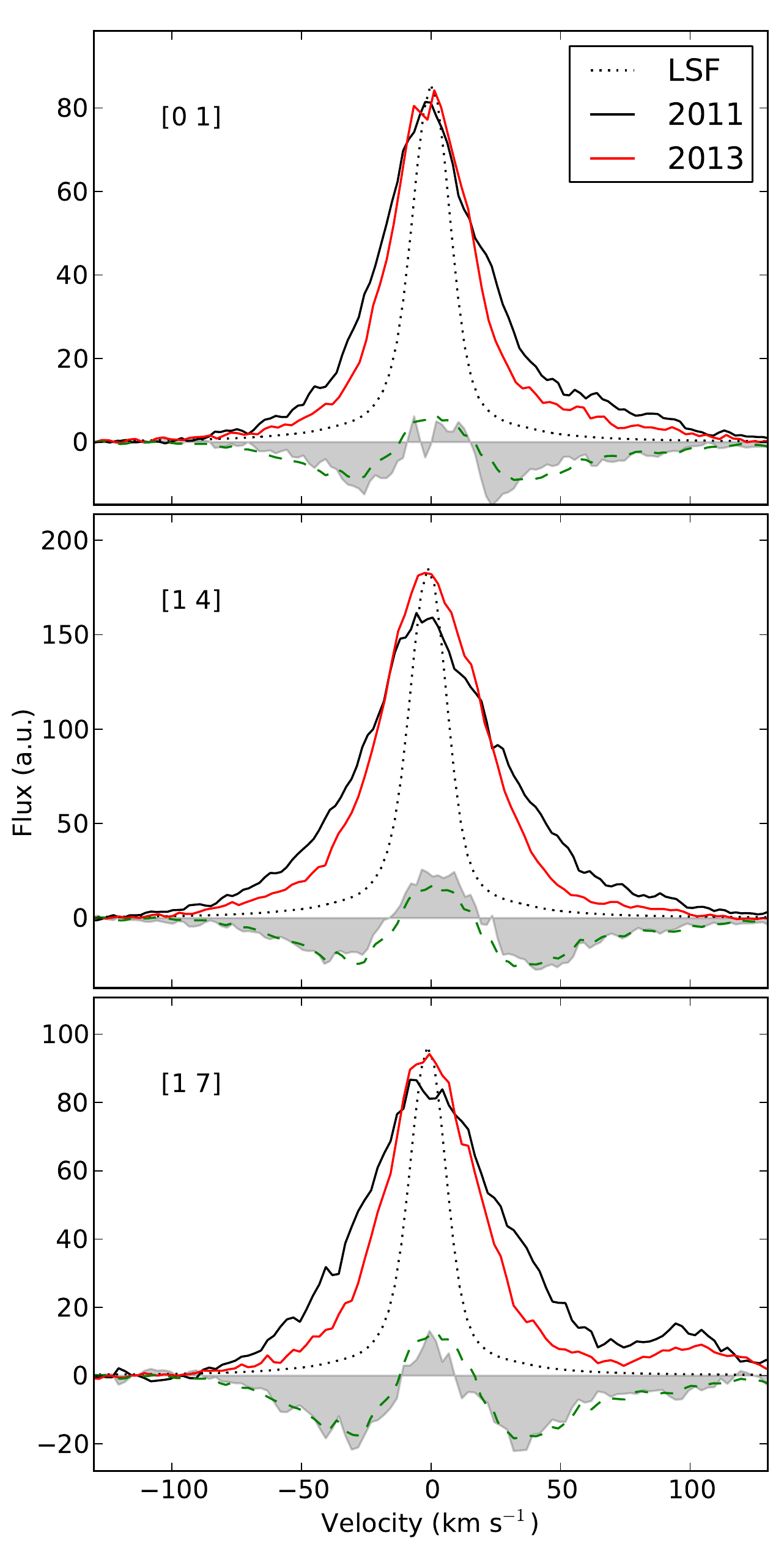}
\caption{Flux reduction in the high velocity of the three strongest H$_{2}$ progressions. Several lines of the respective progressions have been co-added (to avoid uncertainties in the wavelength calibration, the lines have been shifted to a common zero velocity before co-addition). The shaded area indicates the difference between the two epochs. The mean difference scaled to fit the individual results is shown as the green, dashed curve. An approximate LSF is also shown. \label{fig:H2_progressions}}
\end{figure}

\subsection{ Evolution of the line profile}
Figure~\ref{fig:H2_progressions} shows the evolution of the three strongest 
H$_2$ progressions and Tab.~\ref{tab:H2_prop} lists the general properties 
of the securely detected fluorescence routes.
All detected H$_{2}$ emission lines are associated with progressions pumped
by Ly$\alpha$.
The average ratio of the H$_{2}$ fluxes ($F_{2013}/F_{2011}$) is $0.76\pm0.07$ 
without significant differences between the various routes, i.e., a uniform decrease in all routes 
(see Fig.~\ref{fig:FWHM} top). 
The H$_2$ profile is approximately symmetric at both epochs. If the disk 
illumination is largely asymmetric, e.g., if the inner warp obscures part 
of the disk from illumination by the central star, no symmetric emission 
profile would be observed
unless the warp was in front of or behind AA~Tau during the two HST COS 
observing epochs. However, such a special 
configuration is unlikely. Allowing a tolerance of $45^\circ$ in phase, 
the statistical chance is only 6\,\% to find the inner warp in front or behind 
AA~Tau during both observations. We therefore conclude that the phase of the 
inner warp does not influence the observed H$_2$ profile significantly.

The H$_2$ flux decrease is smaller than 
the photospheric flux decrease which demonstrates that the majority 
of the H$_2$ emission is not subject to the
additional absorption, i.e., the upper disk layers are
essentially unaffected while the direct sight line toward the star 
is strongly obscured. This strengthens the interpretation that the extra
absorber is located within the disk and not on the outskirts  of the system
or in the ISM.

Figure~\ref{fig:H2_progressions} shows that the observed flux drops mainly at 
higher velocities 
($|v|\gtrsim20$\,km\,s$^{-1}$) while the flux close to the line center
increases slightly or stays constant. 
As a rough estimate, we can translate the velocity of $20\,$km\,s$^{-1}$ into a 
radius of about two au using Kepler's law. 

\subsection{Disk emission modeling of the H$_2$ emission \label{sect:H2_modelling}}
\begin{figure}
\centering
\includegraphics[width=0.5\textwidth, angle=0]{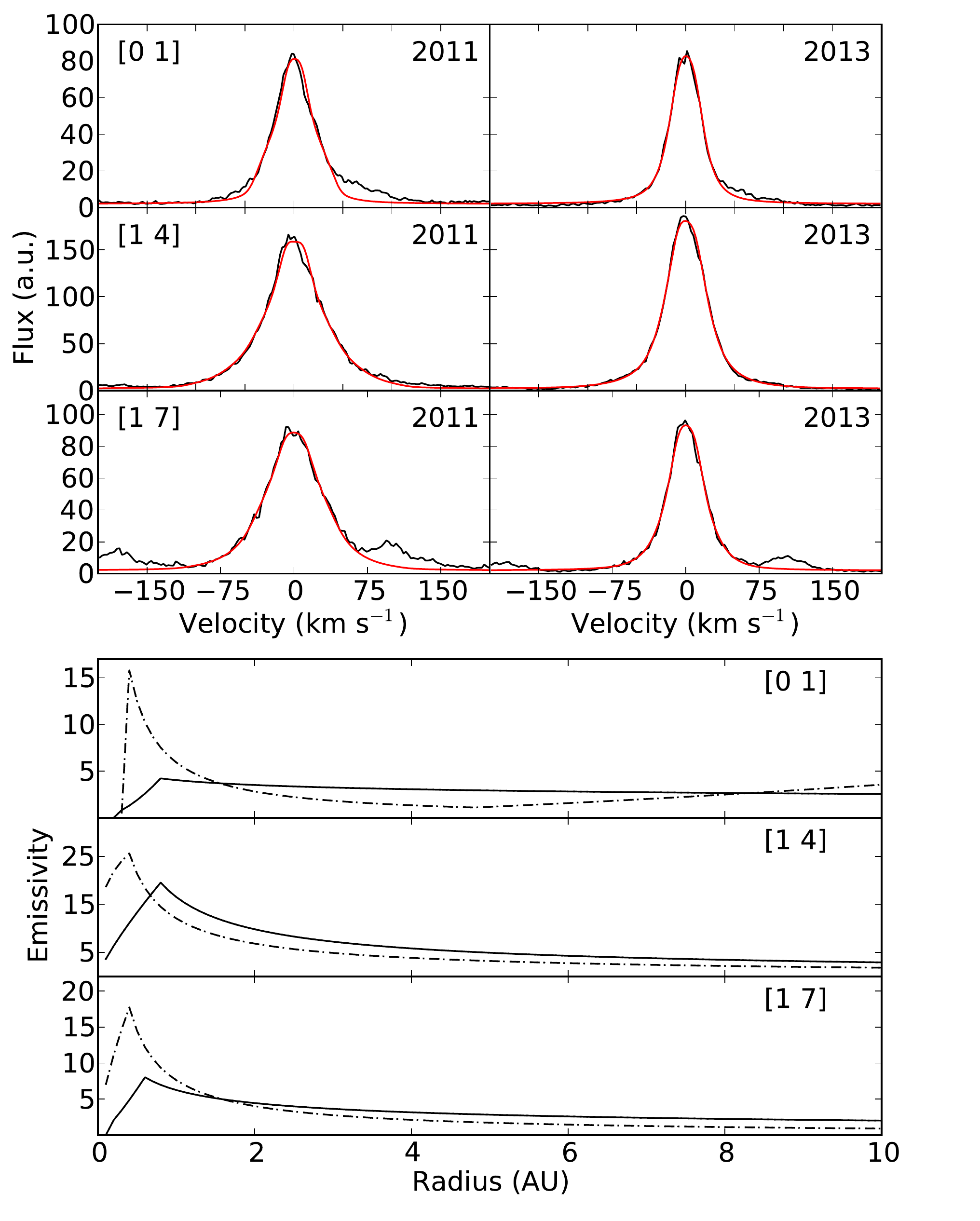}
\caption{Modeled H$_2$ emission lines and emissivity of the 
disk for the different models. {\bf Top}: Data and model for individual
progressions. {\bf Bottom:} Disk emissivity. The 2011 model is shown 
dash-dotted and the 2013  is the solid curve. Details are given in 
sect.~\ref{sect:H2_modelling}. 
\label{fig:H2_models} }
\end{figure}

To improve the above rough estimate for the region suffering extra 
absorption in 2013 compared to 2011, we model the observed H$_2$ profile 
with disk emission profiles. We assume
the broken power law description for the disk emission of 
\citet{Salyk_2011} and describe the emission as
\begin{equation}
L = a \left( \int_{r_{in}}^{r_{mid}} r^p D(r) dr + \frac{r_{mid}^p}{r_{mid}^q}\int_{r_{mid}}^{r_{out}} r^q D(r) dr \right)\,,
\end{equation}
where $D(r)$ is the emission per disk annulus.
We explored the following parameter space: $r_{in}$: 0.1 -- 0.5\,au in 
steps of 0.1\,au, $r_{mid}$: 0.5 -- 5.0\,au in steps of 0.2\,au.
We also experimented with the values for $r_{out}$, but found
little dependence and fixed it to 10\,au.
The parameters $p$, $q$ as well as 
the normalization $a$ were allowed to vary freely.
More sophisticated models might describe the true emissivity
more accurately. However, we are mainly interested 
in the change between 2011 (bright state) and 2013 (dim state)
so that this description is sufficiently accurate for 
our purposes.
Figure~\ref{fig:H2_models} shows the fits and the resulting radial
distribution of the disk emission. 
It shows
that the {\it observed} H$_2$ emission within about one au is 
reduced compared to 2011 while the H$_{2}$ emission coming from 
larger radii remained mainly unchanged with approximately the same 
flux as in 2011.

\subsection{Scenarios for the evolution of the H$_2$ emission}
There are two qualitatively different explanations for the 
reduction of only 
the high velocity H$_2$ emission: (I) absorption of the inner disk emission by 
an outer absorber ($r\gtrsim1\,$au) and (II) a reduction of the H$_2$ 
excitation in the inner disk. For illustrative purposes, we divide model I into
 an absorber at about two au (Ia) and an absorber location at 
larger distances, e.g., 10\,au  (Ib). 
Models (Ia) and (Ib) are intended to represent two realizations of 
an outer absorber that are drawn from a continuum of possible outer
absorber locations. 

\begin{figure*}[th]
\centering
\includegraphics[width=0.89\textwidth, angle=0]{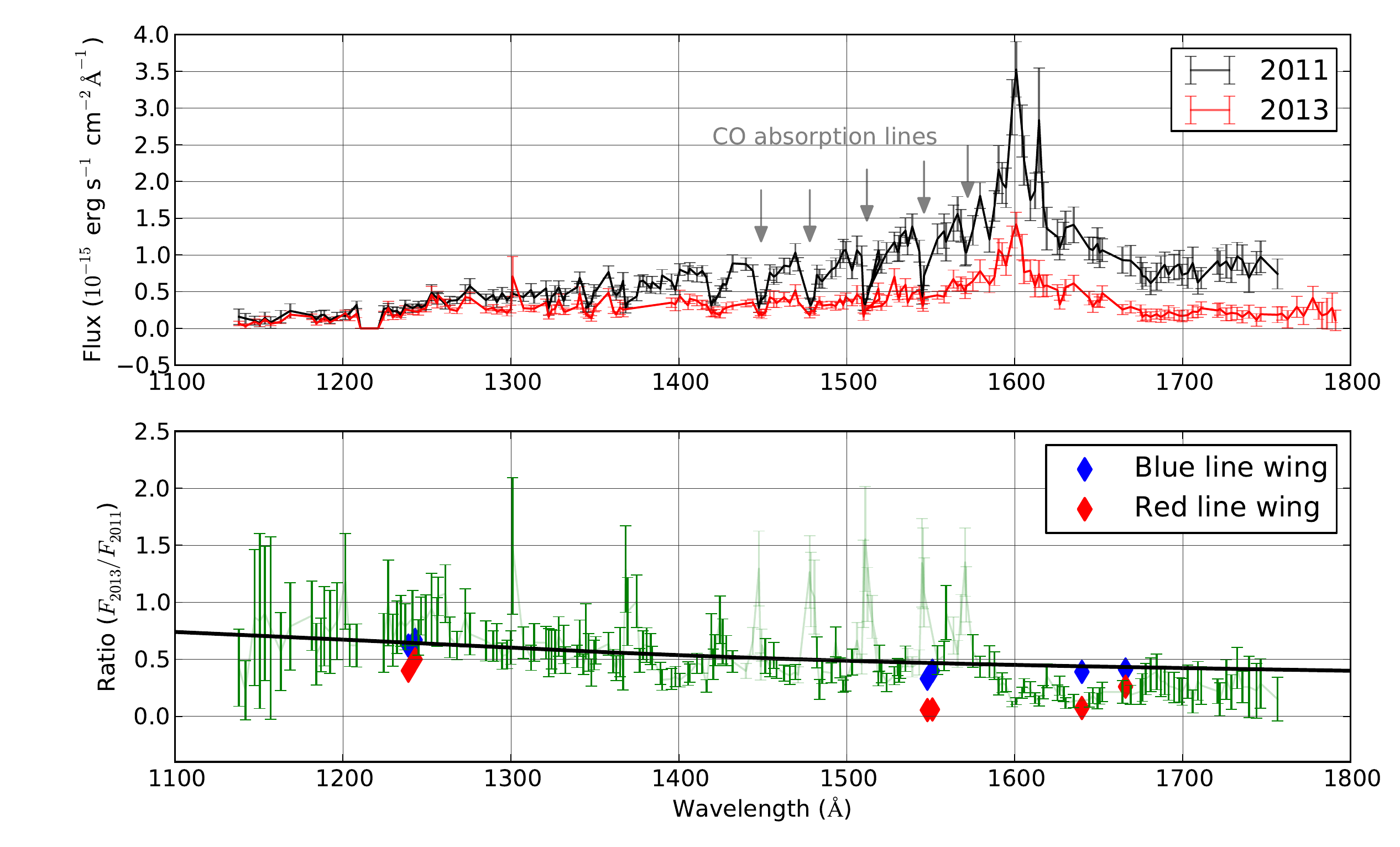}
\caption{{\bf Top panel}: FUV continuum of AA~Tau. The CO absorption lines are visible. The bump at 1600\,\AA{} is also visible. {\bf Bottom panel}: Ratio  between both epochs and the ratio between the two transmission curves discussed in the text.
Overplotted are the blue and red parts of atomic emission lines.\label{fig:cont} }
\end{figure*}

\subsubsection{Scenario I: An outer absorber}
Scenario Ia: Locating the extra absorber at two au would obscure the high 
velocity H$_2$ while leaving the emission from larger radii unchanged.
However, such a 
configuration would also cast a shadow on the parts of the disk behind 
the absorber as viewed from the stellar surface, i.e., part of the outer 
disk. This appears unlikely given that the 
azimuthal extent of the absorber must be $\gtrsim180^\circ$ to remove
about half of the high velocity H$_2$ emission (red and blue shifted emission
are similarly affected). Consequently, a large fraction 
of the disk at $r>2\,$au would not be illuminated with Ly$\alpha$ photons
causing a significant drop in total H$_2$ flux. 
Such a severe drop in the H$_2$ flux would not occur if a similar part of 
the disk was already in the shadow of the absorber in 2011. As the 
H$_2$ profiles are rather symmetric, the absorber should be
located at about the opposing location in 2011 ($180^\circ$ phase shift, 
i.e., mostly behind AA~Tau). The orbital period at two au is about 3.2\,yrs
so that there is sufficient time for the warp to rotate in front
of AA~Tau between the two FUV observations. However, we 
know that the dim state lasts more than four years now, i.e., longer than the 
orbital period at two au. Therefore, a warp at two au would shadow the outer 
disk as seen from the star leaving little to no Ly$\alpha$ illumination to 
pump the H$_2$ emission. Furthermore, assuming a standard height profile
$
h(r) = 0.1 r^{9/7}
$
requires an absorber height of 0.8\,au which is four times the 
value expected  ($h\approx0.1\,r$) and it is challenging to
create a scenario which produces such a massive
disturbance of the disk on the orbital time scale.

Scenario Ib: As a variance of the previous scenario, we now locate the 
absorber at ten au.
In that case, only a stripe  of the disk would be obscured from view while
most regions remain observable from Earth. In addition, the 
Ly$\alpha$ illuminated disk remains unchanged within ten au where
supposedly most of the H$_2$ emission originates.
An azimuthal extent of $20^\circ$ is sufficient to
obscure half of the high velocity H$_2$ emission.
The orbital period at ten au is 35\,yrs so that the extra 
absorber would rotate about $20^\circ$ within the two years 
between the two COS observations, i.e., the angular extent needed to
obscure part of the inner H$_2$ emission.  
Also, the deviation from the 
$h\approx0.1\,r$ relation is reduced compared to the
Ia model (height of 1.8\,au at ten au from AA~Tau) and there 
is no need for a fast dynamic development.
Therefore, we prefer the outer location of the absorber for the cause of
the dimming event and the reduction of the high velocity H$_2$ emission.

\subsubsection{Scenario II: Reduced H$_2$ pumping}
Alternatively, a decrease of the high velocity H$_2$ emission might be caused by
reduced H$_2$ pumping within the inner disk region. An increase of the 
height of the inner disk wall absorbs the stellar Ly$\alpha$ photons
before they can excite H$_2$ molecules in the inner disk while the outer
disk is still illuminated in a flaring disk configuration. However, we consider
this scenario unlikely, because (a) the wall height should be azimuthally symmetric
as the high velocity wings are rather symmetric as well, (b)
the innermost gas disk where the highest velocity H$_2$ resides would not
be affected contrary to the observations, and (c) the increased inner 
disk height would not cause the extra absorption towards AA~Tau as
the viewing angle into the AA~Tau system is 
much larger than the typical angle of the disk height with respect to 
the midplane
($\tan(i\approx15^\circ)=0.27 > 0.1 h/r$ ), i.e, the inner disk
wall would completely shadow the disk if it was in the line of sight.
Lastly, (d) an increase of the inner disk height would be seen
as a flux increase in the K band (or even at shorter wavelengths) which
is not observed (the UKIRT data rather suggests less emission from the inner disk). 
Finally, a scenario where the reduced 
molecular emission is caused by an
inner  disk intercepting less Ly$\alpha$ photons due a lower flaring angle
appears unlikely, because it is also unrelated to the increased absorption.
We therefore prefer an outer absorber as the cause
for the dimming.

\subsection{FUV continuum emission: Grain growth, scattering or a non-stellar origin?}
There is currently no agreement on the origin of the FUV continuum in CTTSs. 
The accretion continuum contributes to the observed flux  (especially at longer 
wavelengths), but a contribution from a yet 
unidentified emission mechanism might dominate the FUV continuum emission
\citep[see discussion in ][]{Ingleby_2009,France_2011b}. 
Given the two FUV observations that experience different amounts of 
circumstellar absorption
towards the stellar source, we investigate the evolution of the continuum
flux to constrain its spatial origin.

Figure~\ref{fig:cont} shows the evolution of the FUV continuum 
\citep[extracted from the COS spectra as described in ][]{France_2014}. 
The continuum flux decreases on average by a factor of about two and 
is bluer in 2013 than in 2011. The FUV continuum flux decrease is 
less than the decrease in optical flux between both epochs. Together 
with the bluer spectral shape,
this shows that just adding absorption cannot explain the observed 
flux evolution. Three qualitatively different possibilities
may explain the flux evolution. First, one can replace part of
the absorber present in 2011 by an absorber with larger grains (i.e.
with $R_V\approx7$) for the 2013 observation; second, scattering
might also contribute to the observed FUV flux; and third, the FUV 
continuum might not originate close to the stellar surface but
within the inner disk region (e.g., at about one au) for which
we know from the H$_2$ flux evolution that the transmission towards
this region decreased by a factor of about two. As outlined in 
App.~\ref{sect:App_cont}, all three possibilities are compatible with available
constraints, however, the grain growth scenario does not explain
the reduction of the optical brightness modulation by the inner warp so 
that scattering or a non-stellar origin of the 
FUV continuum are more likely scenarios. From these two, we favor the
non-stellar origin, because one would naively expect that the stellar emission
lines and the continuum evolve similarly in the scattering scenario
which is not observed. In addition, the flux of the molecular bump at 
1600\,\AA{} decreases similarly as the surrounding continuum suggesting
that both regions experience the same absorption, i.e., originate in
the same region different from the stellar surface.
The scattering and non-stellar origin scenarios would also provide a 
natural explanation 
for the low molecular column densities obtained from CO absorption
against the FUV continuum \citep{McJunkin_2013}.

\section{Summary and discussion \label{sect:summary}}
AA~Tau recently experienced a dimming event caused by circumstellar
material ($\Delta V\sim3$\,mag). We obtained new XMM-Newton X-ray, 
HST FUV  as well as ground-based optical and NIR data of AA~Tau during 
this dim state. 
The analysis of the FUV H$_2$ emission lines shows that the absorption 
is caused by an extra absorber which is most likely located several au 
from AA~Tau, i.e., significantly further out in the system than the 
inner warp (0.1\,au). The decreased NUV to NIR 
fluxes and the reduced modulation of the optical light curve are 
best explained by scattering within the circumstellar environment of 
AA~Tau, as already suggested by \citet{Bouvier_2013}. Since the scattering 
efficiency strongly decreases towards longer wavelengths, 
the NIR data most accurately trace the increase in dust extinction between 
the bright and dim state;  we find
$\Delta A_V\approx4.0_{-0.5}^{+	1.0}$.

Our X-ray data show that the absorbing gas column 
density increases by less than a factor of $<1.6$ 
between the dim and the bright state. 
An increase in absorbing gas  column 
density has just been published by \citet{Zhang_2015}
from an analysis of mid-IR CO absorption lines observed during the dim state;
they find an increase that is higher than 
our X-ray result assuming the canonical CO to H ratio of $10^{-4}$
($N_H^{MIR}>3.2\times10^{22}$\,cm$^{-2}$ vs 
$N_H^{X-ray}<1.5\times10^{22}$\,cm$^{-2}$  and 
$<2.0\times10^{22}$\,cm$^{-2}$ for \citet{Anders_1989} and \citet{aspl} 
abundances, resp.). The higher column 
density is incompatible with the X-ray 
data (see App.~\ref{sect:CO_abso}).
We think that the factor of $>1.5$ difference is caused by a combination
of the following effects. First, the extra absorber might not have 
the canonical CO-to-H ratio of $10^{-4}$ although this has been observed
for RW~Aur \citep{France_2014_CO}. Second, the tension between the 
parameters ($T$ and $b$) for FUV and MIR observed CO absorption indicates
that the CO absorption model contains additional uncertainty; e.g., $T$ and 
$b$ were held fixed during all epochs in the \citet{Zhang_2015} model which
might not be true in this case. 
Third, the chemical composition of the absorber might 
differ from the abundance pattern that we used to convert X-ray absorption
to hydrogen column density.  Appendix~\ref{sect:CO_abso} shows that
the 
discrepancy between $N_H^{MIR}$ and $N_H^{X-ray}$ reduces for lower 
absorber metallicities but remains significant. 
Fourth, the phase of the inner disk warp might affect the observed CO 
column density so that the increase in $N_H$ caused by the extra absorber 
might be lower than the  value derived by \citet{Zhang_2015} .

By comparing X-ray (gas) and optical extinction (dust), we find that the
gas-to-dust ratio of the extra absorber is close to the ISM value.
This contrasts with previous X-ray data of the AA~Tau system that showed 
a gas-to-dust ratio elevated by about a factor of ten with respect to 
the ISM for the 
inner part of the system ($\lesssim0.1$\,au) as suggested 
by previous studies \citep[][]{Schmitt_2007, Grosso_2007}.
Therefore, the new AA~Tau X-ray 
data indicate that the excess gas extinction is restricted to the
inner region around the star while the gas-to-dust ratio is 
within a factor of two to three from the ISM value further out in 
the system, i.e., gas and dust are well mixed in the part of the 
disk traced by the sightline through the extra absorber. 

The appearance of the extra absorber within a $\sim$year time scale
calls for an azimuthally asymmetric disturbance of the 
disk height that rotated into view.
Azimuthal asymmetries are often seen in
scattered light or thermal dust emission, e.g., LkH$\alpha$~330 
\citep[][]{Isella_2013}, but
the spatial scales probed by these 
imaging studies are typically larger than 
the sizes considered here.
The presence of the extra absorber in AA~Tau suggests that
such asymmetries extend to or at least exist also within about 
ten au. 

In general, X-ray observations indicate enhanced gas-to-dust ratios for
CTTSs \citep[see, e.g., Tab.~4 in][ for a recent comparison between dust
and X-ray extinction for a number of CTTSs]{McJunkin_2013b}.
Without the detailed monitoring of the AA~Tau system, one could not 
make the distinction between gas-rich inner region and ISM-like outer
region and would divide the total gas column density by the
total dust extinction resulting in gas-to-dust that is about
twice the ISM-ratio, i.e., the omnipresent gas-rich absorber.
However, if the AA~Tau result  with a gas-rich inner region around
the star and a rather ISM-like gas-to-dust ratio further out 
is representative for CTTSs in general, it 
suggests that the excess gas absorption is caused by material
confined to the innermost region ($\lesssim0.1$\,au), i.e.,
accretion funnels or a wind launched close to the star.
Such an absorber consisting mostly of gas can significantly reduce the 
amount of high energy photons impacting the protoplanetary disk 
without being visible in optical observations. This might have severe 
consequences for the disk 
chemistry which strongly depends on the irradiation.

\begin{acknowledgements}
It is a pleasure to thank N. Schartel for granting the target of opportunity
XMM-Newton observation, S. Wolk for organizing the UKIRT observation, B. Reipurt 
for providing the UKIRT data, and the CAHA staff for performing the NIR observations
in service mode; in particular Aceituno-Castro, Bergond, and
the director Galad\'i-Enr\'iquez for granting the target of opportunity 
observations. 
PCS was supported by the DLR under grant 50 OR 1307 and is currently a 
Research Fellow at ESA/ESTEC. KF acknowledges from a Nancy Grace Roman 
Fellowship and HMG acknowledges NASA/Chandra Award GO4-15009X
issued by the Chandra X-ray Observatory Center for and on behalf of NASA
under contract NAS8-03060. 
We acknowledge LabEx OSUG@2020 that allowed purchasing the imaging 
system installed at the 1.25-m telescope at CrAO.
Based on observations obtained with XMM-Newton, 
an ESA science mission with instruments and contributions directly funded 
by ESA Member States and NASA. The paper is also based on observations 
obtained by the Hubble Space Telescope through Guest Observing Program 12876. 
Some of the data presented in this paper were obtained from the Multimission 
Archive at the Space Telescope Science Institute (MAST). STScI is operated 
by the Association of Universities for Research in Astronomy, Inc., under NASA 
contract NAS5-26555. Support for MAST for non-HST data is provided by the NASA 
Office of Space Science via grant NAG5-7584 and by other grants and 
contracts. We acknowledge with thanks the variable star observations from 
the AAVSO International Database contributed by observers worldwide and 
used in this research.

\end{acknowledgements}

\bibliographystyle{aa}
\bibliography{aa_tau}
\appendix

\section{Optical magnitudes during the HST observations \label{sect:HST_mags}}
The evolution of the optical magnitude between the two HST observations
is required to check if particular absorber properties are reasonable,
e.g., if they provide a sufficient amount of V band extinction.

For the 2011 HST observation, we use the STIS acquisition images obtained just
prior to the COS observations 
to estimate the optical brightness assuming an effective temperature of 4000\,K
and $A_V=0.78$ as found by \citet{Bouvier_2013} for the bright state. We
find a V-magnitude of \hbox{about 13.5} which is consistent with a short HST 
STIS G430L spectrum preceding the COS exposures (the G430L partly overlaps
with the V bandpass). This magnitude is typical for the bright state 
of AA~Tau and indicates some absorption by the inner disk warp, because the
brightness of the uneclipsed state has been about 12.5 in the V-band.

For the 2013 HST observation, no preceding optical acquisition images nor optical
spectra are available so that we estimate the V-magnitude from contemporary 
optical data points\footnote{
    \texttt{http://vizier.u-strasbg.fr/local/Vbin/\\\hspace*{5.0cm}VizieR?-source=35570077}}
\citep{Bouvier_2013}.
They indicate 
$V\approx15\,$mag, i.e., about 1.5 magnitudes lower than during the previous
COS observations.

\section{Period analysis}
\begin{figure}[t!]
\centering
\includegraphics[width=0.48\textwidth, angle=0]{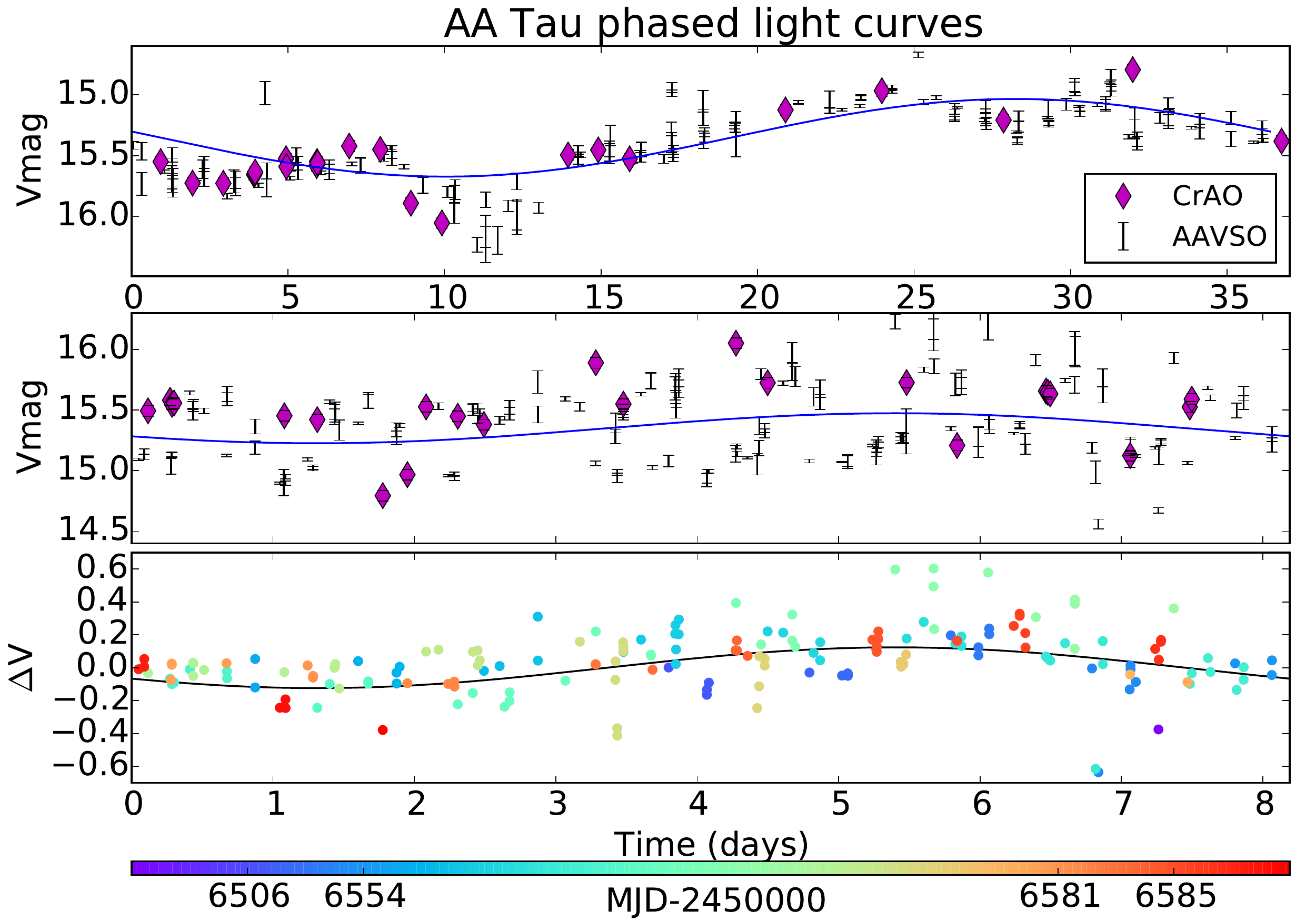}
\caption{Phased optical light curves. The color coding in the bottom panel
indicates the observing time as encoded in the colorbar.\label{fig:timing_new} }
\end{figure}

Figure~\ref{fig:timing_new} shows phased optical light curves of AA~Tau.

\section{FUV continuum evolution \label{sect:App_cont}}
The broad wavelength coverage of the HST COS spectra 
allows us to determine the wavelength dependent evolution of 
the FUV continuum flux which, in the simplest scenario, constrains
the dust population of the extra absorber.
However, additional processes like grain growth, scattering 
and a non-stellar origin of the FUV continuum hamper direct
conclusions and, in fact, none of these three mechanisms can
be strictly ruled out. 

\subsection{Grain growth}
Adding extra extinction cannot cause the continuum to appear
bluer in 2013 than in 2011. However, replacing part of the
absorber present in 2011 with an absorber that contains on average
larger grains causes a bluer continuum.
Specifically, replacing ISM-like absorption
with $A_V=0.4$\,mag  by an absorber with $A_V=1.9$ and $R_V=7$ in the 
\citet{CCM} description provides a reasonable description
of the flux evolution and is compatible with the evolution
of the V magnitudes between both epochs (see App.~\ref{sect:HST_mags} 
for the evolution of the V magnitudes between both epochs and 
Fig.~\ref{fig:cont} (bottom) for the expected continuum flux 
evolution).

As we have a better data coverage for the time around the
XMM-Newton observation, we checked if adding further 0.6\,mag of
optical extinction with $R_V=7$ in the \citet{CCM} description provides
a reasonable description of the flux evolution between the 
bright state and the dim state around 
the XMM-Newton observation and found that this, again, requires the brighter 
NIR fluxes contrary to the observations
(see discussion of the 
grain growth scenario in sect.~\ref{sect:scattering}).
It is impossible to bring the
lower NIR fluxes  in agreement with the XMM-Newton optical/NUV fluxes by
replacing part of an ISM-like absorber during the bright state 
with an absorber with $R_V>3.1$ for the 2013 observation. Thus,  
scattering in the
optical is still required to achieve agreement with the lower NIR fluxes when replacing part
of the 2011 absorber with an $R_V>3.1$ one in 2013. 
Explaining the FUV continuum evolution (moderate flux decrease and blueing) 
with a modification of the absorber therefore requires scattering in
the optical but not in the FUV. In addition, this scenario
requires a different explanation for the evolution of the red shifted atomic FUV emission
lines which decrease more strongly than the surrounding continuum.

\subsection{Scattering}
Scattering can easily explain the blueing of
the continuum by postulating that the shortest wavelength already contained 
scattered photons in 2011 while this did not apply to the same extent to
the longer wavelengths. Removing (part of) the direct emission then 
causes a flux drop as well as a bluer appearance of the continuum. 
This
is compatible with scattering in the optical, but does not explain that
the red shifted atomic emission drops more strongly.

\subsection{A non-stellar origin}
The analysis of the H$_2$ emission lines showed that the
increased absorption mainly affects the line of sight towards
the star while, e.g., emission from the outer disk remained 
largely unaffected.
In particular, the high velocity H$_2$ emission reduces by about the
same fraction as the FUV continuum suggesting that the FUV continuum 
originates also within 
the inner few au of the disk. However, the blueing of the continuum 
remains unexplained in this scenario; potentially scattering also
contributes to the continuum even if its origin is not close to the
stellar surface or the shape of the non-stellar emission depends
on the distance to the star with redder emission predominately emitted
at smaller radii. This is compatible with the evolution of the emission bump 
at 1600\,\AA{} seen in both epochs. This bump
has been interpreted in terms of a H$_2$ molecular dissociation quasi-continuum
generated by collisions with non-thermal electrons 
\citep[e.g.,][]{Herczeg_2004, Ingleby_2009}, i.e., clearly non-stellar. 
Its flux reduced similarly as the surrounding continuum suggesting that 
the bump and the majority of the continuum probably experience similar 
extinction and, thus, probably have a similar spatial origin, too.

To summarize, a non-stellar origin of the FUV continuum is compatible
with the observational constrains and scattering might contribute as well
while the grain growth scenario appears less likely.

\section{FUV emission lines \label{sect:app_FUV_lines}}
The atomic FUV emission lines are usually assumed to be emitted
close to the stellar surface and, thus, should experience the
same absorption as the photospheric emission.
Furthermore, the effect of extinction is more pronounced in the FUV than in 
the optical. Therefore, we investigate the evolution of the atomic FUV
emission lines and concentrate our analysis on the strongest lines 
(C~{\sc iv}\,$\lambda\lambda\,1548,\,1551$, 
N~{\sc v}\,$\lambda\lambda\,1238,\,1242$ and He~{\sc ii}$\,\lambda\,1640$).

\subsection{Details of the data analysis}
We removed the 
H$_{2}$ ``contamination'' from the atomic emission lines
using the approach presented by \citet{France_2014}, i.e., by 
fitting the wavelength region under consideration with a number of 
Gaussians convolved with the instrumental line profile and removing 
components identified as strong H$_{2}$ fluorescence lines.
Lastly, we split the lines into a blue and red shifted part,
because shock heated gas within the jet can increase 
the observed blue shifted emission
\citep[e.g., C~{\sc iv} in 
the DG~Tau jet, see][]{Schneider_2013a}.
Further, we exclude velocities with $|v| < 50\,$km\,s$^{-1}$ as the
COS wavelength calibration is not accurate to within a few 
10\,km\,s$^{-1}$.

COS does not observe these lines
simultaneously so that time variability during the observations 
can influence the results when comparing data from different
wavelength settings. Indeed, there is minor
variability during the 2011 observations with a slight drop of the 
C~{\sc iv} and N~{\sc v} count-rates over several hours which 
could change the C~{\sc iv} to N~{\sc v} ratio by about 20\,\% 
assuming a continuous, linear evolution. In 2013,  the rates remain 
relatively constant ($<10\,\%$ changes). In the following we neglect 
these effects, because the variability between the two epochs 
is much larger.

\subsection{Results}
\begin{figure}[th!]
\centering
\includegraphics[width=0.49\textwidth, angle=0]{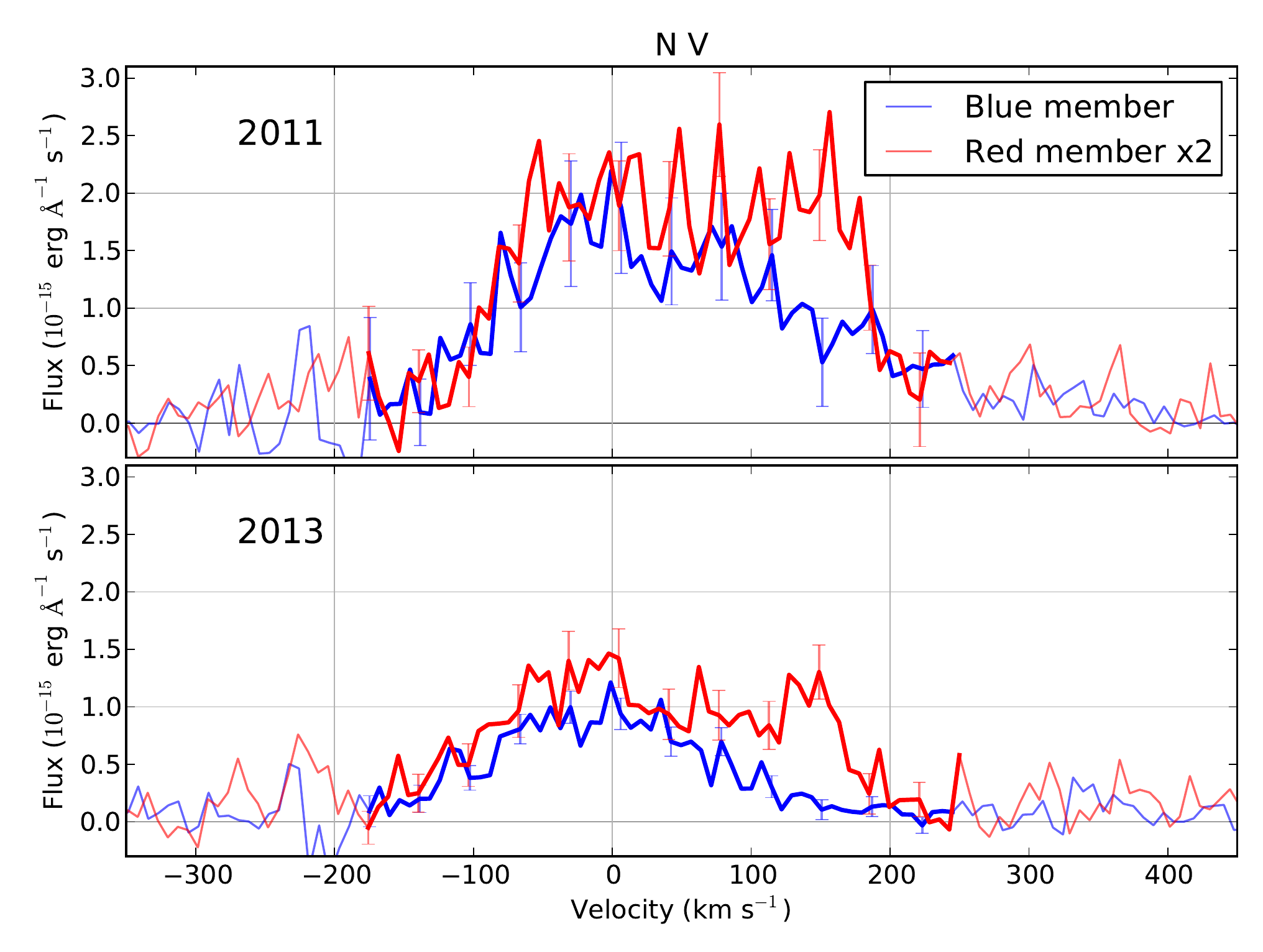}
\includegraphics[width=0.49\textwidth, angle=0]{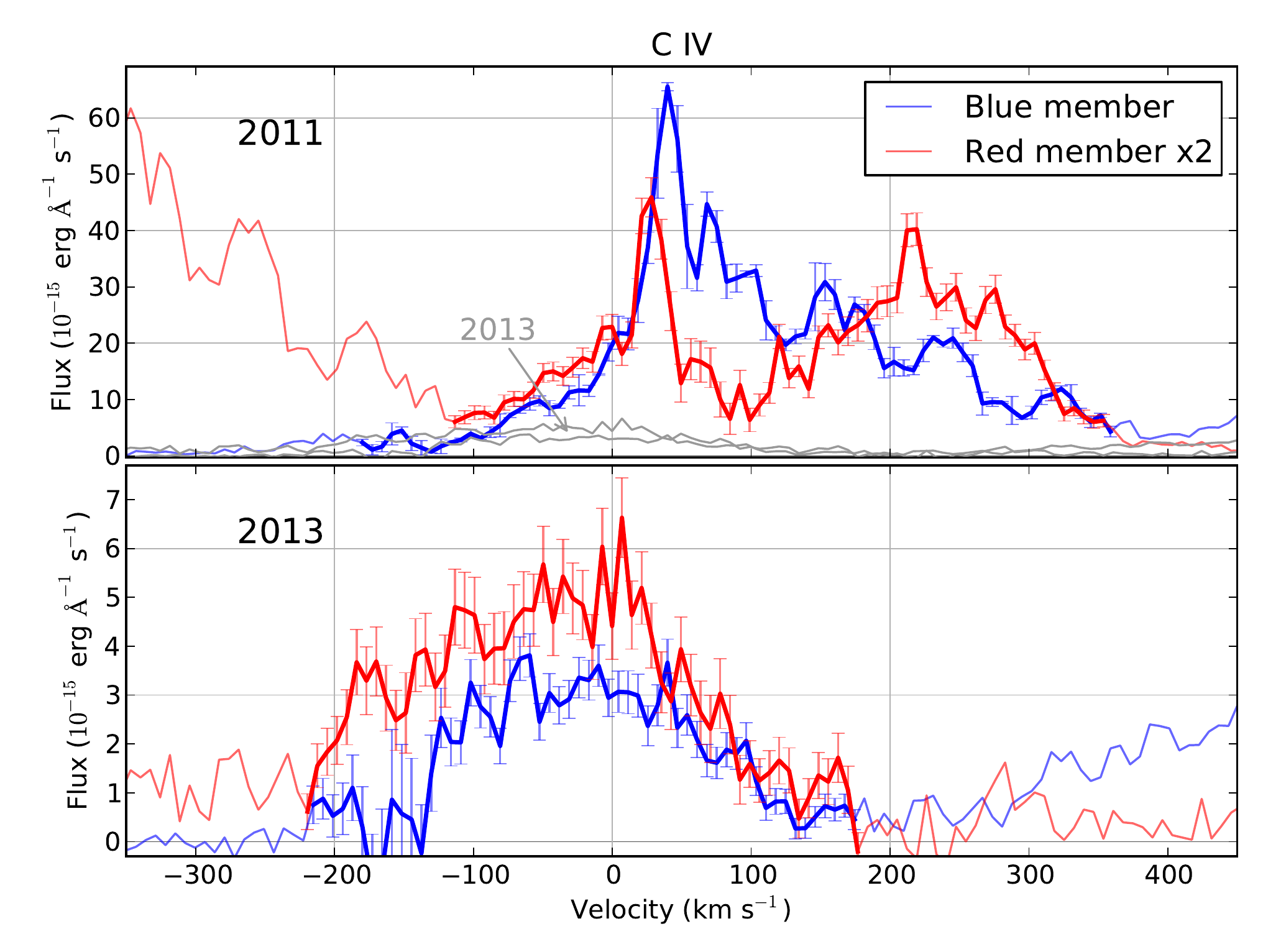}
\includegraphics[width=0.49\textwidth, angle=0]{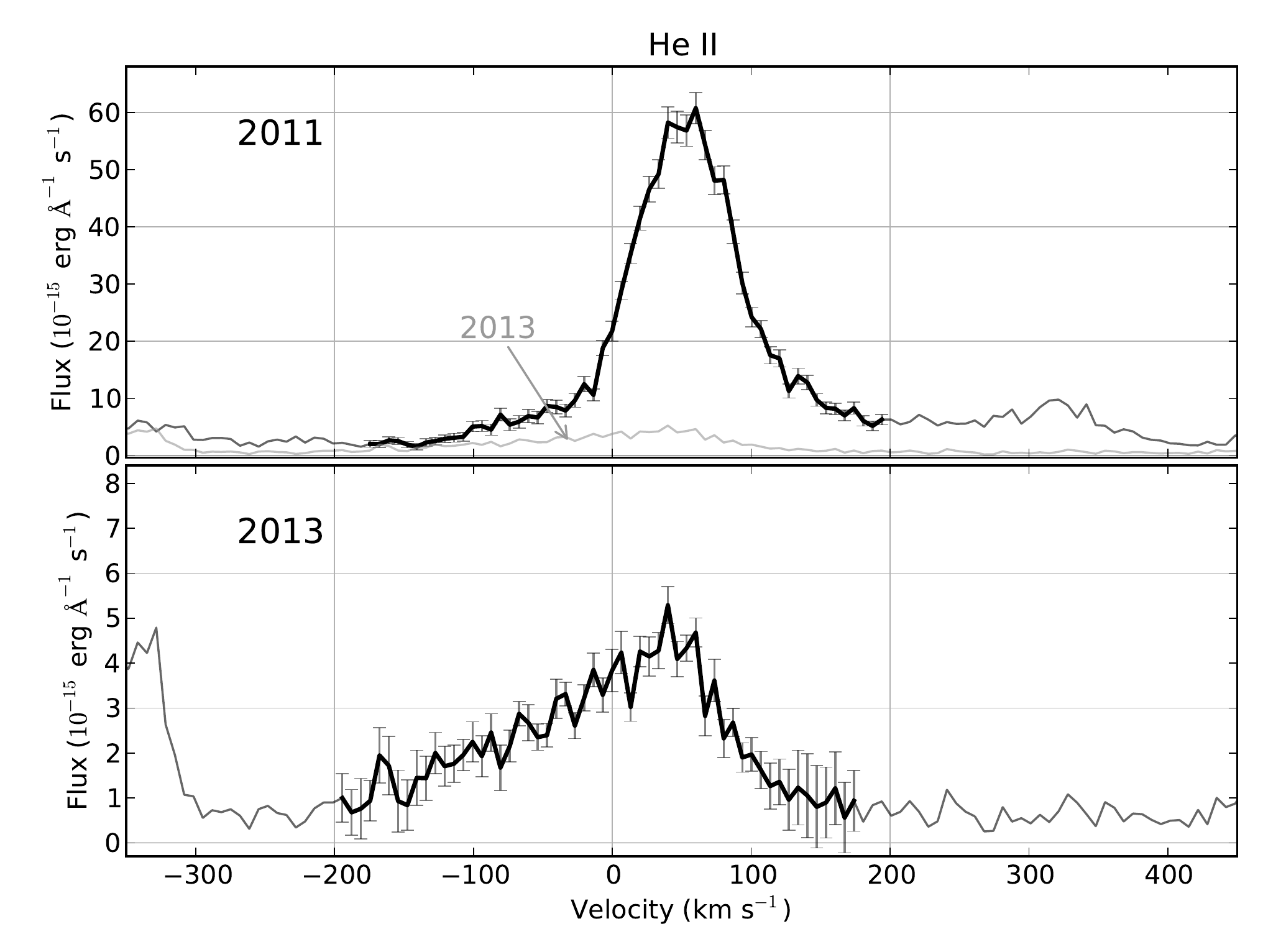}
\caption{Comparison of the blue and red members of the N~{\sc v} and 
C~{\sc iv} doublets as well as the evolution of the He~{\sc ii} line. 
The red members have been scaled by a factor of two to easily see the 
optical depth effects. The relevant parts of the lines are drawn with 
thick lines and errors. \label{fig:dc} }
\end{figure}

Figure~\ref{fig:dc} shows the atomic emission lines and their
line fluxes are listed in Tab.~\ref{tab:afluxes} (which also includes 
the properties of the  C~{\sc ii} and O~{\sc iii}] lines).
The red and blue shifted emission evolve differently between
both epochs with the red shifted part reducing more strongly 
than the blue shifted one. 
This might be due to the fact that
red shifted emission should be dominated by emission
related to the accretion process (accretion funnels or spots)
while the blue shifted emission contains a higher fraction of jet/outflow 
emission which originates above the disk and is, thus, not subject
to the extra absorption.
For both red and blue shifted emission, the real flux drop might 
be higher than observed due to scattered photons.
In addition, the red line wing might be 
``contaminated'' by broad, slightly blue-shifted emission lines from the outflow/jet.
Furthermore, the atomic emission lines in the FUV are time variable
so that tight constrains on the properties of the absorber 
cannot be derived. With these caveats
in mind, we describe the evolution of the red and blue shifted
emission separately in the following.

\subsubsection{Red shifted emission}

\begin{table*}[t]
\begin{minipage}[h]{0.99\textwidth}
\centering
\renewcommand{\footnoterule}{}
  \caption{Line fluxes for selected atomic lines and continuum\label{tab:afluxes} }
  \begin{tabular}{c c c c r r r} \hline \hline
  Line-ID & Rest wavelength  &  Velocity & Flux 2011 & Flux 2013 & Ratio\\     
          & ($\AA$) & range (km\,s$^{-1}$) & ($10^{-15}$\,erg\,s$^{-1}$\,cm$^{-2}$) & ($10^{-15}$\,erg\,s$^{-1}$\,cm$^{-2}$) & \\
  \hline
  N {\sc v}  & 1238.8 & -130 \dots +240 &  $1.71\pm0.05$ & $0.77\pm0.02$ & $0.45\pm0.02$\\
             & 1242.8 & -130 \dots +240    &  $1.16\pm0.04$ & $0.65\pm0.02$ & $0.56\pm0.03$\\
             & sum & blue wing & $0.79\pm0.04$ & $0.58\pm0.02$ & $0.73\pm0.05$\\
             &     & red wing & $1.78\pm0.05$ & $0.6\pm0.02$ & $0.34\pm0.02$\\
  C {\sc ii} & 1335.7  & total\footnote{0\dots200 for 2013, 10\dots250 for 2013} & $2.46\pm0.05$ & $1.30\pm0.03$  & $0.53\pm0.02$\\
  C {\sc iv} & 1548.2 & -300 \dots 150    &  $32.53\pm0.38$ & $3.56\pm0.16$ & $0.110\pm0.005$\\
             &        & total\footnote{$-300\dots+400$ for 2011, $-300\dots+150$ for 2013} & $50.09\pm0.45$ & $3.56\pm0.16$ & $0.071\pm0.003$\\
             & 1550.8 & -80 \dots 400    &  $22.50\pm0.34$ & $2.36\pm0.09$ & $0.105\pm0.004$\\
             &        & total\footnote{$-80\dots+400$ for 2011, $-300\dots+350$ for 2013}          & $22.50\pm0.34$ & $3.74\pm0.10$ & $0.166\pm0.005$\\
             & sum & blue wing & $7.1\pm0.2$ & $3.5\pm0.2$ & $0.49\pm0.03$\\
             &     &  red wing & $57.3\pm0.5$ & $2.1\pm0.1$ & $0.04\pm0.01$\\
  O {\sc iii}] & 1666.2 & total\footnote{$-300\dots540$ for 2011, $-300 \dots 150$ for 2013}              & $2.76\pm0.20$ & $0.85\pm0.08$ & $0.30\pm0.04$\\
             &        & -300 \dots +150 & $1.75\pm0.16$ & $0.85\pm0.08$ & $0.49\pm0.06$\\
              & & blue wing & $0.72\pm0.10$ & $0.43\pm0.05$ & $0.59\pm0.11$\\
              & & red wing & $0.97\pm0.10$ & $0.24\pm0.05$ & $0.24\pm0.06$\\
  He {\sc ii} & 1640.4 & -240 \dots 200 & $32.11\pm0.41$ & $3.00\pm0.15$ & $0.093\pm0.005$\\
              & & blue wing & $2.09\pm0.16$ & $1.05\pm0.09$ & $0.50\pm0.06$\\
              & & red wing & $23.94\pm0.34$ & $1.32\pm0.10$ & $0.06\pm0.01$\\[0.4cm]
  \multicolumn{2}{c}{Continuum} & & Average Flux 2011\footnote{Numbers in brackets give integrated fluxes in units of $10^{-14}$\,erg\,s$^{-1}$\,cm$^{-2}$} & Average Flux 2013 & Ratio \\
   \multicolumn{2}{c}{Wavelength range} && ($10^{-15}$\,erg\,s$^{-1}$\,cm$^{-2}$\,\AA)  & ($10^{-15}$\,erg\,s$^{-1}$\,cm$^{-2}$\,\AA) &\\
   \hline
  \multicolumn{2}{c}{$\lambda < 1300$\,\AA{}} & & $0.26\pm0.01$ (4.0) & $0.19\pm0.01$ (3.0)&  $0.73\pm0.05$\\
  \multicolumn{2}{c}{$1300 < \lambda < 1600$\,\AA{}} & & $0.79\pm0.01$ (24.4) & $0.39\pm0.01$ (10.4) & $0.49\pm0.01$ \\
  \multicolumn{2}{c}{$\lambda > 1600$\,\AA{}} & & $0.90\pm0.01$ (42.1) & $0.37\pm0.01$ (16.5) &  $0.41\pm0.01$\\
  \hline
  \end{tabular}
  \end{minipage}
\end{table*}

The red shifted wing of all atomic emission lines show lower fluxes 
during the 2013 observation compared to 2011. The fluxes in the red 
wings of C~{\sc iv} and He~{\sc ii} decrease much more strongly than 
the N~{\sc v} flux (see sum/red wing in Tab.~\ref{tab:afluxes})
and  more strongly than the FUV continuum.
The different evolution of C~{\sc iv} and N~{\sc v} fluxes is unexpected,
because both species trace rather similar plasma temperatures 
($\log T\sim 5.0$ and 5.3) and are usually well correlated \citep{Yang_2012,
Ardila_2013}. 
The N~{\sc v} to C~{\sc iv} ratio is closer to the CTTS correlation \citep{Ardila_2013}
in 2013 than in 2011. This suggests that the observed N~{\sc v} flux was
unusually low in 2011. Because the N~{\sc v} flux was already low 
in 2011, its factional decrease was less than for the other FUV emission
lines. We therefore concentrate on C~{\sc iv} and He~{\sc ii} in 
the following.

\begin{table}[t]
\renewcommand{\footnoterule}{}
\caption{Extinction values ($A_V$) for red shifted FUV emission lines. The \citet{CCM} description of the extinction has been used for the $R_V=7$ column and ``Special'' indicates that an ISM-like absorber with $A_V=0.4$ has been 
replaced by an $R_V=7$ absorber with $A_V$ as given in the Table. \label{tab:FUV_extinct} }
\begin{minipage}[h]{0.49\textwidth}
  \begin{tabular}{c c c c c}\hline\hline
  Line & Obs. ratio & ISM-like & $R_V=7.0$ & Special \\
  \hline
  N~{\sc v}    &  $0.34\pm0.02$ & 0.4 & 1.3 & 2.2 \\
  C~{\sc iv}   &  $0.04\pm0.01$ & 1.4 & 3.6 & 4.5 \\
  He~{\sc ii}  &  $0.06\pm0.01$ & 1.2 & 3.1 & 4.1 \\
  \hline
  \end{tabular}
\end{minipage}  
\end{table}

Tab.~\ref{tab:FUV_extinct} summarizes the absorption
required to explain the flux evolution for different descriptions of
the FUV extinction assuming no intrinsic variability. It shows that 
the evolution is compatible with
ISM-like absorption based on the evolution of the V magnitude, 
but incompatible with the evolution of the NIR magnitudes which, 
unfortunately, have not been obtained simultaneously.
The evolution is also compatible with an additional $R_V=7$ 
absorber. In addition, it is possible to replace part of 
an ISM-like absorber in 2011 by an $R_V=7$ absorber 
assuming  scattering contributes to the observed optical 
but not to the FUV fluxes since this is still compatible with the 
range of observed NIR magnitudes. 
Given potential intrinsic time variability, the FUV emission 
lines are compatible with the different absorber properties
discussed above. Furthermore, they are even compatible with 
higher $A_V$ values for an ISM-like absorber assuming that
scattering operates also on the atomic FUV emission lines
or emission in the red shifted wing of a blue shifted emission
component contributes to the red shifted velocity range.

\subsection{Blue shifted atomic emission}
The evolution of the blue 
line wings of the atomic emission lines is visualized in Fig.~\ref{fig:cont}.
The blue wings decrease similarly as the surrounding FUV continuum.
This suggests that both components are similarly affected by the
extra absorber, i.e., less strongly than the red shifted emission
originating close to the stellar surface.

Interpreting the continuum emission in terms of a non-stellar origin easily 
reconciles the evolution of the blue shifted atomic emission lines as they 
probably originate partly in the outflow/jet, 
i.e., above the disk. In particular, the C~{\sc iv} emission in 2013 appears
slightly blue shifted and rather Gaussian shaped as expected for an outflow/jet
origin. Whether the observed atomic emission traces knots within the jet
as in DG~Tau \citep{Schneider_2013a} or a hot stellar wind remains undecided.
The similar evolution of blue shifted atomic emission lines and
continuum favors a similar spatial origin.
Knots within the 
jet further away from the source would not be affected by the extra absorber, 
so that the an origin close to the disk is likely which favors a stellar 
wind as the origin of the blue shifted emission lines.

\section{CO emission \label{sect:CO_emiss}}
Fluorescent CO emission traces the coolest FUV-emitting plasma observed in
our COS spectra. With a typical temperature of a few hundred K it is 
significantly colder than the observed H$_2$ emission 
\citep[about $2-3\times10^3$\,K, ][]{France_2011b}.
The flux of the strong CO emission lines is constant indicating
that the CO emitting region is not significantly
affected by the extra absorber and that 
the pumping radiation field did not change significantly 
between the two FUV observations.

\begin{figure}
\centering
\includegraphics[width=0.5\textwidth, angle=0]{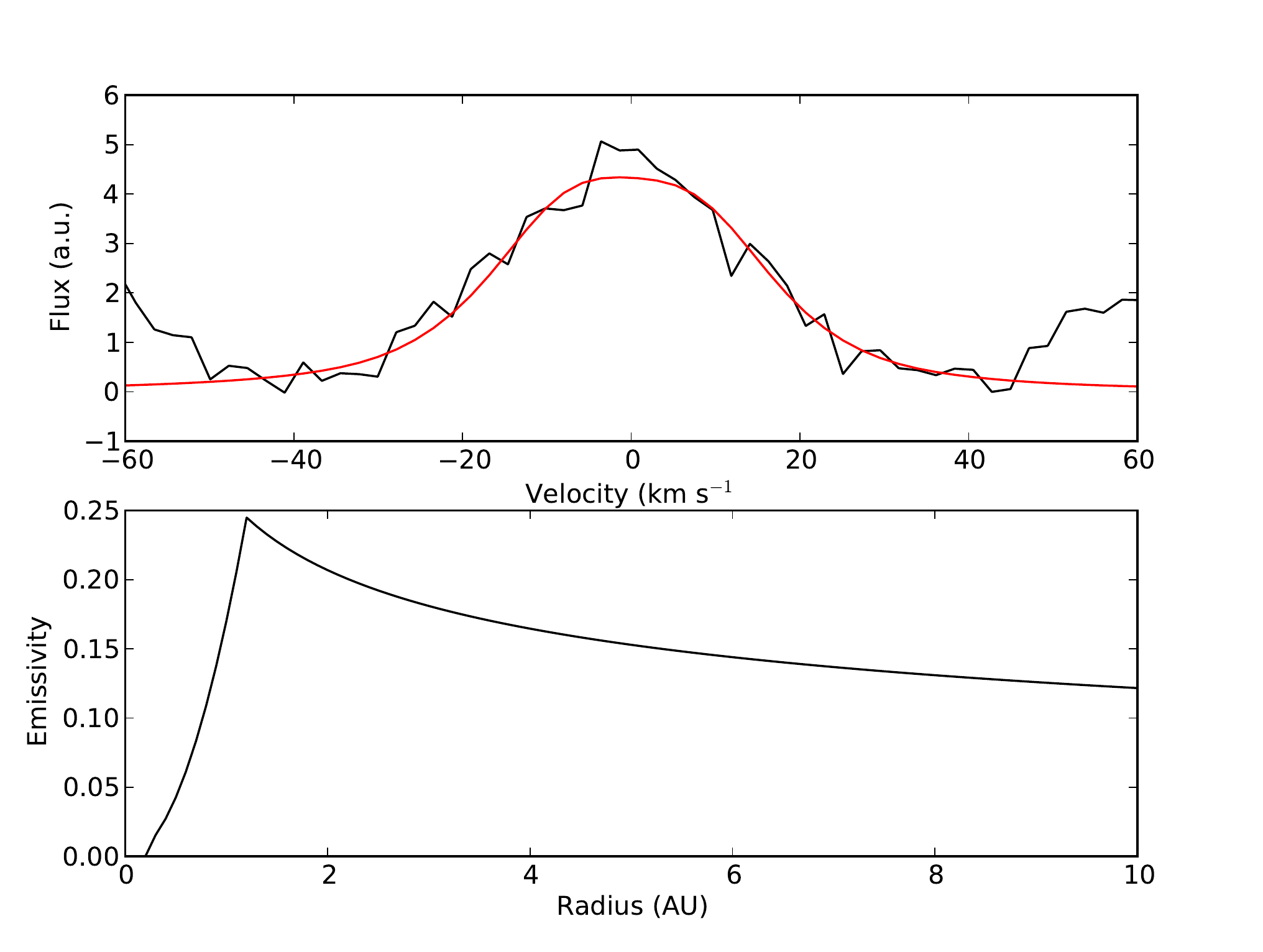}
\caption{{\bf Top}: Fit to the 2013 CO emission. {\bf Bottom}: Emissivity of the 
disk. \label{fig:CO_models} }
\end{figure}

In principle, we could also apply the disk emission modeling 
to the CO emission to constrain the spatial origin of the emission.
Unfortunately, this is only possible with some accuracy for the CO 
data from 2013 but not for 2011. Figure~\ref{fig:CO_models} shows that
the CO disk emissivity peaks at slightly larger radii than the H$_2$
disk emissivity (CO: 1.2\,au, H$_2$:
0.6-0.8\,au), but appears in general rather similar to the H$_2$
models for 2013. Therefore, the CO emission comes from similar
regions within the disk as the H$_2$ in 2013. Since the total flux did
not change significantly, a similar spatial origin in 2011 is likely.
This again indicates that only the disk emission within one to two au is
affected by the extra absorber.

\begin{figure}
\centering
\includegraphics[width=0.5\textwidth, angle=0]{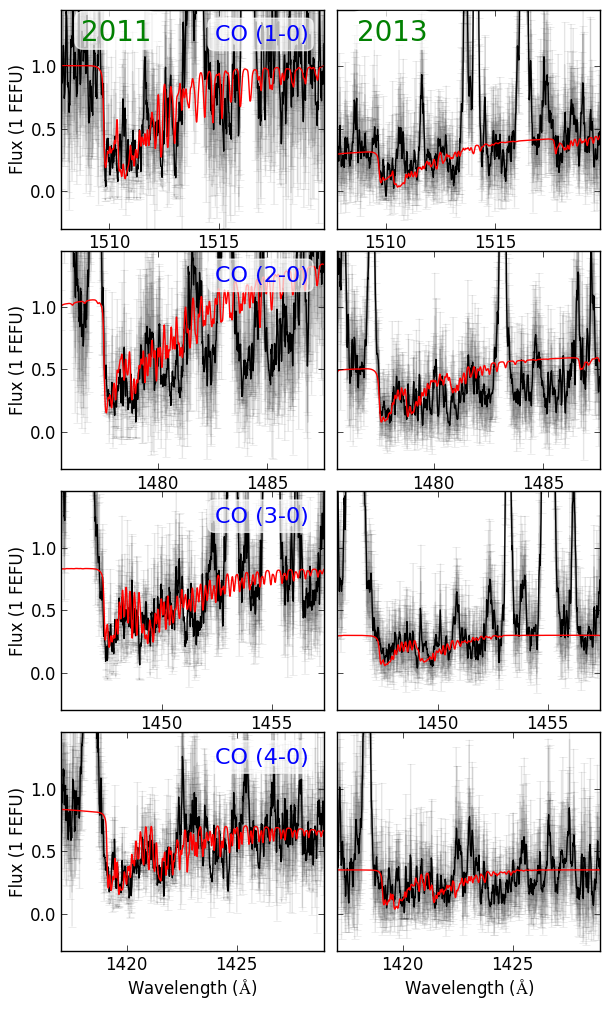}
\caption{CO absorption bands with fits. \label{fig:CO_abso} }
\end{figure}

\section{CO absorption \label{sect:CO_abso}}
Figure~\ref{fig:CO_abso} shows the evolution of the CO absorption seen
in COS. To compare the CO absorption between both epochs, we fit the continuum
data (see sect.~\ref{sect:App_cont}) around the CO absorption bands with a first
order polynomial excluding the data falling into the wavelengths ranges of
the 1-0 to 4-0 absorption bands. Using this continuum, we calculate an 
``equivalent width'' (EW) of the CO absorption by summing the wavelengths 
regions unaffected by H$_2$ emission and find that the  EW decreases by about
10--21\,\% between both epochs ($1\,\sigma$ confidence range). Including the 
uncertainty in the continuum level which we assume to be about 10 and 15\,\% 
for the 2011 and 2013 data, the EW ratio between both epochs is  $1.16\pm0.40$.
Thus, the CO absorption column densities are similar unless the Doppler 
broadening parameter $b$ differs significantly between both epochs. 
Unfortunately, $b$ is not well constrained due to the large instrumental 
line width (17\,km\,s$^{-1}$) compared to the $b$ value which is usually 
around $1$\,km\,s$^{-1}$ \citep{McJunkin_2013}. Nevertheless, we attempted
to fit  the CO absorption using the methods of \citet{McJunkin_2013} and 
find a best fit CO column density of $2.5\times10^{18}\,$cm$^{-2}$, a 
temperature of 100\,K and a Doppler $b$ value of 0.5 km s$^{-1}$ for the 
2013 data. This column density is about an order of magnitude higher than 
that derived by \citet{France_2012_AA_Tau} for the 2011 observation, which is 
mainly caused by the lower $b$ value (0.5 vs 1.2 km\,s$^{-1}$), but also 
by the much cooler temperature (100 vs. 500\,K). However, the errors on 
the CO column densities are about 1\,dex. In summary, we find no significant
increase of the CO absorption column density.  

While working on the revised version of this paper, \citet{Zhang_2015} published
CO mid-IR data of AA~Tau during the dim state. They find 
$N_{^{12}CO}=3.2\times10^{18}$\,cm$^{-2}$, which is close to our best fit value 
from the FUV CO absorption. However, their temperature and broadening values
($T\approx494$\,K and 2.2\,km\,s$^{-1}$, resp.) differ significantly from our
absorption model. Adopting their higher broadening value would strongly 
decrease the column density required to explain the FUV CO absorption. We thus
conclude that both measurements might not trace the same absorber. Given that
scattering might contribute to the observed FUV emission, one possibility is
that the FUV absorption traces CO located at higher disk altitudes than the 
MIR CO absorption which is unaffected by scattering.

Using the canonical CO to H ratio of $10^{-4}$, \citet{Zhang_2015} estimate 
$N_H>3.2\times10^{22}$\,cm$^{-2}$ for the extra absorber, which is of the 
same order of magnitude as our column density derived from the X-ray data.
However, their lower limit is incompatible with the X-ray data. 
Using the minimum column density absorbed in 2003 
($N_H=0.9\times10^{22}$\,cm$^{-2}$) plus 
their minimum column density ($N_H=3.2\times10^{22}$\,cm$^{-2}$)
for the extra absorber results in $N_H>4.1\times10^{22}\,$cm$^{-2}$, but
models with $N_H=4.1\times10^{22}$\,cm$^{-2}$ do not provide an 
acceptable fit to the X-ray data. They result in $\chi^2=86$ (51 d.o.f.) 
using the \citet{Anders_1989} abundances and 
in $\chi^2=78$ using the \citet{aspl} abundances, respectively.
The best fitting model has $\chi^2=33 / 34.5$ (\citet{Anders_1989}, \citet{aspl} abundances,
resp.). The 90\,\% confidence
range upper limit on the absorbing column density from the X-ray data is 
$2.3\times10^{22}$\,cm$^{-2}$ \citep{Anders_1989} and 
$3.3\times10^{22}$\,cm$^{-2}$ \citep{aspl}.

\section{Additional H$_2$ information}
\begin{figure}
\centering
\includegraphics[width=0.49\textwidth, angle=0]{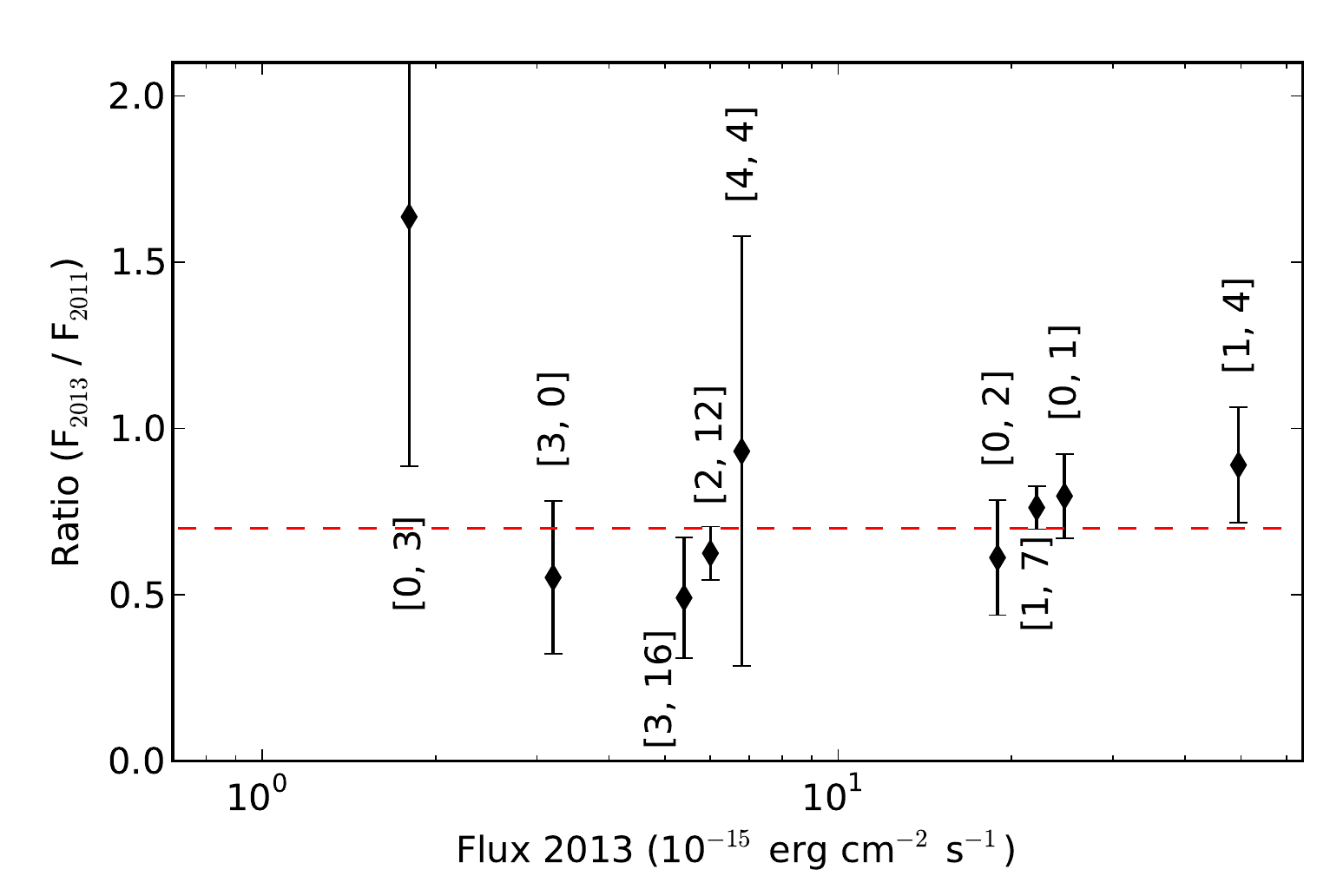}
\includegraphics[width=0.49\textwidth, angle=0]{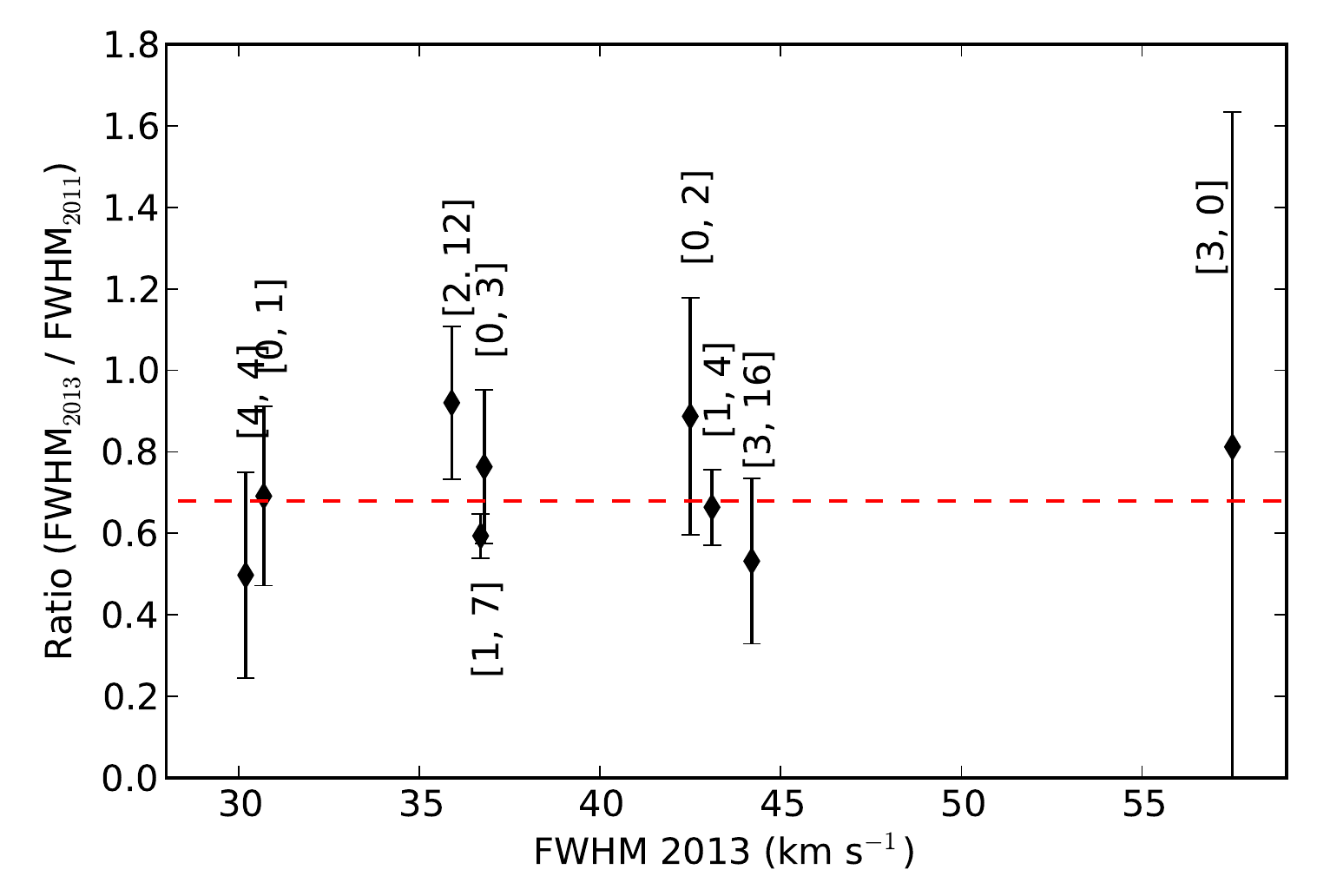}
\caption{Evolution of the H$_2$ flux and FWHM for different progressions.\label{fig:FWHM} }
\end{figure}

Table~\ref{tab:H2_prop} lists the properties
of the detected H$_2$ progressions and Fig.~\ref{fig:FWHM}
shows the evolution of the fluxes and width of the
H$_2$ lines.
 
\begin{table}[t]
\setlength{\tabcolsep}{4pt}
\begin{minipage}[h]{0.49\textwidth}
\centering
\renewcommand{\footnoterule}{}
\caption{Properties of the molecular hydrogen progressions during the two epochs.
\label{tab:H2_prop}}
\begin{tabular}{c | r r | r r} 
\hline \hline
 & \multicolumn{2}{c}{FWHM (km s$^{-1}$)} & \multicolumn{2}{c}{ Flux (10$^{-14}$ erg cm$^{-2}$ s$^{-1}$ )}\\ 
$[v^{'}$, $J^{'}]$  & 2011 & 2013 & 2011 & 2013 \\
\hline
$[0,1]$ & 44.4~$\pm$~13.9   &   30.7~$\pm$~1.7   &        31.0~$\pm$~4.8  &  24.7~$\pm$~0.8 \\  
$[1,4]$ & 64.9~$\pm$~6.6   &   43.1~$\pm$~4.1   &         55.6~$\pm$~9.1  &  49.5~$\pm$~5.3 \\
$[1,7]$ & 61.8~$\pm$~3.8   &   36.7~$\pm$~2.5   &       29.0~$\pm$~1.8  &  22.1~$\pm$~1.3 \\
$[2,12]$ & 39.0~$\pm$~4.2   &   35.9~$\pm$~6.2   &     9.6~$\pm$~0.5  &  6.0~$\pm$~0.7 \\
$[0,3]$  & 48.2~$\pm$~7.4   &   36.8~$\pm$~7.1   &      1.1~$\pm$~0.5  &  1.8~$\pm$~0.1 \\
$[0,2]$ & 47.9~$\pm$~11.3   &   42.5~$\pm$~9.7   &       30.9~$\pm$~5.8  & 18.9~$\pm$~4.0 \\
$[3,0]\footnotemark[1]$ \footnotetext{$^1$ Low-quality 2011 fits, comparison probably not meaningful} & 70.8~$\pm$~40.2   &   57.5~$\pm$~48.2   &      5.8~$\pm$~1.6  &   3.2~$\pm$~1.0 \\
$[4,4]\footnotemark[1]$ & 60.7~$\pm$~24.7   &   30.2~$\pm$~9.2   &      7.3~$\pm$~4.8  &   6.8~$\pm$~1.5 \\
$[4,13]\footnotemark[1]$ & 78.1~$\pm$~N/A   &   36.7~$\pm$~N/A   &     17.6~$\pm$~N/A  &   12.9~$\pm$~N/A \\ 
\hline
\end{tabular}
\end{minipage}
\end{table}

\end{document}